\documentclass[twocolumn,aps,amsmath,superscriptaddress,longbibliography,notitlepage]{revtex4-2}

\usepackage{amssymb}
\usepackage{mathrsfs}
\usepackage{graphicx}
\usepackage{float}
\usepackage[caption=false]{subfig}
\usepackage[colorlinks=true,linkcolor=blue,citecolor=blue]{hyperref}

\usepackage{verbatim}
\usepackage{enumitem}
\usepackage{multirow}
\usepackage{lipsum}
\usepackage[dvipsnames]{xcolor}
\usepackage{hyperref}
\usepackage{bm}
\usepackage{physics}
\usepackage{makecell}

\usepackage[ruled,vlined]{algorithm2e}

\def\blue{\textcolor{blue}}
\def\Rev{\textcolor{black}}

\definecolor{amber}{rgb}{1.0, 0.49, 0.0}

\newcommand{\ro}[1]{{\color{black} #1}}

\begin{document}

\title{Designing non-Hermitian real spectra through electrostatics}

\author{Russell Yang} 
	\email{russell.yang@chch.ox.ac.uk}
	\affiliation{Department of Physics, National University of Singapore, Singapore 117551, Republic of Singapore}
	\affiliation{Clarendon Laboratory, University of Oxford, Oxford OX1 3PU, UK}
	\author{Jun Wei Tan}
	\affiliation{Department of Physics, National University of Singapore, Singapore 117551, Republic of Singapore}
	\author{Tommy Tai}
	\affiliation{Cavendish Laboratory, University of Cambridge, Cambridge CB3 0HE, UK}
	\author{Jin~Ming~Koh}
\affiliation{Division of Physics, Mathematics and Astronomy, Caltech, Pasadena, CA 91125, United States}
\author{Linhu Li}  
\affiliation{Guangdong Provincial Key Laboratory of Quantum Metrology and Sensing $\&$ School of Physics and Astronomy, Sun Yat-Sen University (Zhuhai Campus), Zhuhai 519082, China}
\author{\ro{Stefano} Longhi} 
\affiliation{Dipartimento di Fisica, Politecnico di Milano, Piazza Leonardo da Vinci 32, I-20133 Milan, Italy}
\affiliation{IFISC (UIB-CSIC), Instituto de Fisica Interdisciplinary Sistemas Complejos - Palma de Mallorca, E-07122 Spain}
	\author{Ching Hua Lee}  \email{phylch@nus.edu.sg}
	\affiliation{Department of Physics, National University of Singapore, Singapore 117551, Republic of Singapore}

\begin{abstract}
	Non-hermiticity presents a vast newly opened territory that harbors new physics and applications such as lasing and sensing. However, only non-Hermitian systems with real eigenenergies are stable, and great efforts have been devoted in designing them through enforcing parity-time (PT) symmetry. In this work, we exploit a lesser-known dynamical mechanism for enforcing real-spectra, and develop a comprehensive and versatile approach for designing new classes of parent Hamiltonians with real spectra. Our design approach is based on a new electrostatics analogy for modified non-Hermitian bulk-boundary correspondence, where electrostatic charge corresponds to density of states and electric fields correspond to complex spectral flow. As such, Hamiltonians of any desired spectra and state localization profile can be reverse-engineered, particularly those without any guiding symmetry principles. By recasting the diagonalization of non-Hermitian Hamiltonians as a Poisson boundary value problem, our electrostatics analogy also transcends the gain/loss-induced compounding of floating-point errors in traditional numerical methods, thereby allowing access to far larger system sizes.\\
	
\end{abstract}
	
\date{\today}

\maketitle		
\section{Introduction}
	Condensed matter physics has traditionally been studied in the Hermitian context, since real energies are necessary for observing stable quantum states. Yet, with intense recent research in non-Hermitian systems~\cite{PhysRevLett.89.270401}, it has become apparent that many of the most exciting contemporary phenomena -- exceptional points~\cite{PhysRevX.6.021007,Hodaei2017,Zhen2015}, non-Hermitian skin localization and modified bulk-boundary correspondence~\cite{PhysRevLett.121.086803,PhysRevB.99.201103,song2019non,PhysRevLett.123.066404,tai2022zoology,zhang2022universal,qin2022non}, nontrivial spectral topology~\cite{Li2021,PhysRevX.9.041015,PhysRevLett.125.126402,PhysRevLett.124.086801,Slager2020,jiang2022filling,liang2022anomalous}, negative entanglement entropy~\cite{chang2020entanglement,lee2020exceptional}, effective non-Hermitian curved spaces~\cite{lv2021curving}, amplified Rabi frequencies~\cite{Lee2020} -- exist only in the non-Hermitian realm. Fortunately, non-Hermitian systems are not necessarily unstable, since they can still possess real eigenenergies if appropriately designed. To guarantee real spectra, a key approach has been to enforce parity-time (PT) symmetry~\cite{bender1998real,ElGanainy2018,feng2017non,PhysRevLett.126.215302}, such that the gains and losses conspire to give rise to eigenstates with conserved total amplitude. \Rev{However, it should be emphasized that having a PT symmetric Hamiltonian is neither a necessary nor a sufficient condition for obtaining real spectra - not sufficient because we also require the PT symmetry to be unbroken~\cite{Mostafazadeh2002}. This additional condition} restricts candidate non-Hermitian systems to particular optical media~\cite{Longhi2017,El-Ganainy2019,lin2017line,Ruter2010} or lattice configurations~\cite{Li2019,Regensburger2012}, thereby forgoing potentially rich possibilities afforded by many other platforms~\cite{ghatak2020observation,Helbig2020,xiao2020non,zou2021observation}. \Rev{On the other hand, PT symmetry is also not a necessary condition because pseudo-hermiticity provides a more definitive condition for real spectra~\cite{doi:10.1063/1.1418246,PMID:30312068}. However, pseudo-Hermiticity exists in our context only in the a priori unknown engineered real spectrum system, and not the parent PBC Hamiltonian to be found. 
	}
	
\begin{figure}
		\includegraphics[width=0.99\linewidth]{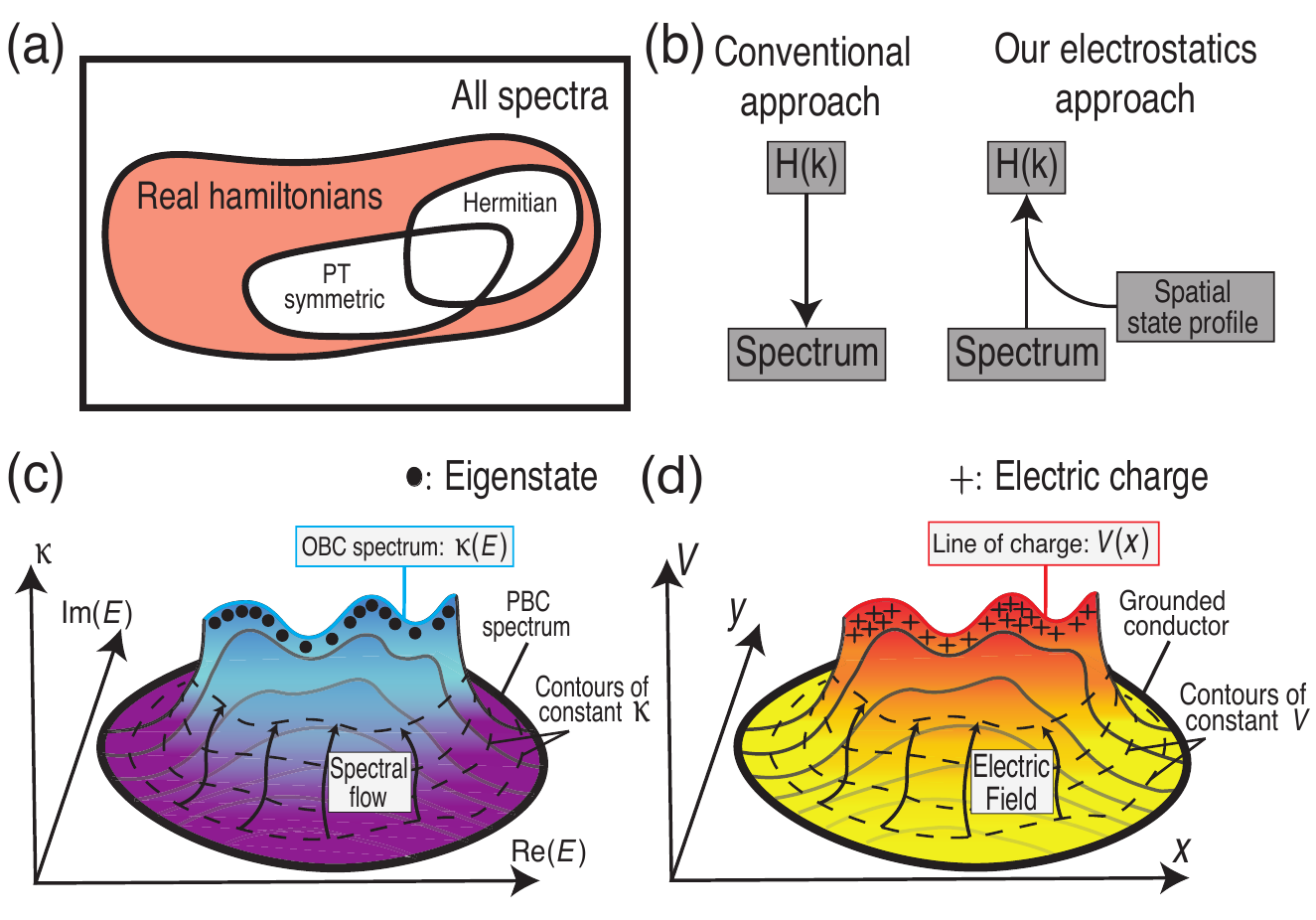}
		\caption{
\textbf{(a)} Hermiticity and PT-symmetry are two well-known routes towards real energy spectra, but our work provides a new approach for designing generic real-spectrum Hamiltonians (pink) satisfying neither condition. \textbf{(b)} Conventionally, model Hamiltonian parameters have to be repeatedly optimized to yield the desired spectrum. By contrast, our electrostatics design approach directly outputs a parent Hamiltonian $H$ possessing almost any desired eigenspectrum and eigenstate profiles. \textbf{(c)} Non-Hermitian eigenstates are characterized by their complex energies $E$ and inverse spatial decay lengths $\kappa$, which together describe a landscape $\kappa(E)$. In particular, PBC eigenstates lie along $\kappa=0$ loops, while OBC eigenstates accumulate along ridges where $\kappa(E)$ is not smooth. \textbf{(d)} The $\kappa(E)$ landscape of a non-Hermitian system is mathematically equivalent to the $V(x,y)$ potential landscape of its electrostatics analog, with PBC and OBC spectral loci corresponding to grounded conductors ($V=0)$ and lines of induced charges respectively. See Tables~\ref{table1}, Sect.~\ref{electroderivation} and the {Supplementary Sections I, II, III, and IV} for more elaboration.}
		\label{fig:1}
	\end{figure}

	\Rev{All in all}, this work is motivated by the lesser-known observation that even without PT symmetry, non-Hermitian gain/loss can still \textit{dynamically} cancel in a bounded medium, if they enter through directed unbalanced couplings. In such cases, states moving in one direction will be amplified, while those moving in another direction will be attenuated. For a bounded system, these gain/loss processes cannot continue forever, and states become stable once they are ``trapped'' by a boundary. This mechanism thus constitutes an alternative approach to real energy spectra (pink region in Fig.~\ref{fig:1}a), in addition to PT symmetry and/or hermiticity. Indeed, such real spectra have been found in simple bounded non-Hermitian lattices with asymmetric nearest-neighbor couplings~\cite{hatano1996localization,LonghiPRA2016}. However, generalizing this dynamical mechanism for real spectra to richer, more complicated lattices remains a challenge, since
	unbalanced couplings also destroy the cherished Bloch property associated with ordinary translation-invariant states, necessitating the additional knowledge of the generalized Brillouin zone (GBZ)~\cite{PhysRevLett.125.226402,PhysRevLett.123.066404}. 
	With the form of the GBZ being difficult to obtain analytically in all but the simplest cases~\cite{PhysRevB.102.085151,PhysRevLett.123.066404,PhysRevLett.125.226402,li2019geometric,wu2021connections}, the comprehensive design of systems with real non-Hermitian spectra thus hinges on a fundamentally more illuminating approach for treating the effects of unbalanced non-Hermitian couplings.
	
	In this work, we circumvent these difficulties by developing a new electrostatics analogy that intuitively reconstructs parent Hamiltonians with desired spectra and spatial eigenstate profiles (Fig.~\ref{fig:1}b), different from known electrostatics analogies that related the correlations within complex non-Hermitian spectra with a Coulomb gas. This contrasts with conventional approaches where the Hamiltonian is fixed through a combination of symmetry arguments and empirical optimization, and the spectrum (desired or otherwise) subsequently computed from it. Being a reversal of conventional approaches, our electrostatics analogy also avoids directly diagonalizing non-Hermitian Hamiltonians, which usually suffers from fatal floating-point errors in large systems~\cite{PhysRevLett.125.226402}.
	
	Our electrostatics analogy exploits the hitherto unexploited parallel between the conformal structure of electric field lines in real space, and the complex spectral flow as non-Hermitian states accumulate along boundaries (Fig.~\ref{fig:1} and Table~\ref{table1}). It identifies non-Hermitian periodic/open boundary condition (PBC/OBC) spectra  respectively with the loci of grounded conductors and electrostatic charges. Their electrical potential distribution gives the corresponding extent of boundary state accumulation, a key property that stabilizes state evolution, with no Hermitian analog. As such, instead of having to solve complicated algebraic equations to determine the GBZ, one just needs to solve the equivalent electrostatics problem, \textit{i.e.} the Poisson equation~\cite{krutitskii2000dirichlet}, which is more geometrically intuitive. In particular, given a set of available couplings, all possible non-Hermitian models with real OBC spectra can be obtained by solving the corresponding electrostatics problem with charges restricted to a line, and grounded conductors determined by the coupling constraints.
	
In the rest of this work, we first explain the mechanism of obtaining real spectra from unbalanced (asymmetric) couplings, and use that to develop our electrostatics analogy and design approach. Next, after two pedagogical warm-up examples with well-known models, we demonstrate how we can robustly discover new models with real spectra, even when unexpected from symmetry arguments. Finally, we discuss our electrostatics reconstruction beyond the context of real spectra, where it is equally mathematically valid, and still physically relevant either as an optimization avenue, or in the context of network Laplacians.

	\section{Results}
	\subsection{Unbalanced couplings, real spectra and their electrostatics analogy}
	\label{electroderivation}

Before deriving our electrostatics analogy, we first recap how unbalanced non-Hermitian couplings can lead to real spectra and hence stable eigenstates. Non-Hermitian systems that possess real spectra experience nontrivial gain and loss, but their effects balance out such that states do not grow or decay indefinitely, whilst retaining novel non-Hermitian properties. Unlike the conventional route of attaining this balance through certain global symmetries, \textit{i.e.} PT symmetry, we shall consider a less known \textit{dynamical} balance approach via unbalanced couplings between lattice sites in a bounded lattice. Away from boundaries (more generally, spatial inhomogeneities like disorder or impurities), unbalanced couplings cause states to grow or decay unimpeded while moving in different directions, resulting in eigenenergies with negative/positive imaginary parts. However, since amplification/decay is tied to spatial motion, further amplification of wavepackets is inevitably resisted upon reaching a boundary, resulting in boundary-accumulated skin states. If this resistance to amplification is perfect, the spectrum will become real.	

The simplest example is given by the Hatano-Nelson model~\cite{hatano1996localization}  
	\begin{equation}
		H_\text{HN}=\sum_j t_+c_j^\dagger c_{j+1} +t_- c^\dagger_{j+1}c_j
		\label{HN}
	\end{equation}
	with unequal nearest neighbor couplings $|t_+|\neq|t_-|$ only. Without boundaries(PBCs), amplification is unimpeded and its PBC spectrum is complex, given by $E_\text{HN}=t_+e^{-ik}+t_-e^{ik}$, $k\in[0,2\pi)$. However, in the presence of a boundary, its OBC spectrum $\bar E_\text{HN}=2\sqrt{t_+t_-}\cos k$ is real~\cite{PhysRevB.102.085151} due to non-Bloch boundary accumulation $\psi(x)\sim \psi(x)\left({t_+}/{t_-}\right)^x$ which induces $k\rightarrow k+i\log\sqrt{\frac{t_-}{t_+}}$. 
	While it is straightforward to understand the perfect balance of gain/loss in this particular simple model, boundaries induce a very complicated deformation $k\rightarrow p(k)=k+i\kappa(k)$ in most other models i.e. the OBC spectrum $\bar E$ converges towards a complex momentum deformation of the PBC spectrum $E(k)$ i.e. $ \bar E(k) =E(k+i\kappa(k))$~\cite{PhysRevB.99.201103}, as elaborated in the {Supplementary Section I}. 
	As a complex extension of the Bloch momentum, $\kappa(k)$ corresponds to real-space decay, and is known as the inverse skin depth of the $k$-eigenstate.
	Mathematically, $\kappa(k)$ must satisfy the condition that there exist at least two different $k,k'$ such that $\bar E(k)=\bar E(k')$ and $\kappa(k)=\kappa(k')$. Such conditions can be expressed as a rather complicated singular function, which is in general difficult to solve.

Alternatively, 
our electrostatics approach sheds geometric intuition to the rather opaque algebraic problem of finding non-Hermitian OBC spectra, and 
provides a more intuitive and tractable approach for engineering the most generic lattice Hamiltonians with real spectra. To understand how, note that the complex deformation $k\rightarrow p(k)$ can be understood geometrically [Fig.~\ref{fig:1}c]: While the PBC spectrum (solid loop) traces out a loop $E(k)$, $k\in [0,2\pi)$ in the complex $E$ plane, the OBC spectrum (line of crests within the loop) is obtained by the ramping up $|\text{Im}(p)|$ such that the PBC loop ``shrinks'' into its interior until it overlaps with itself i.e is degenerate everywhere~\cite{PhysRevB.99.201103,PhysRevLett.124.086801,zhang2021tidal,li2021non}.  In this OBC limit, the spectrum $\bar E$ traces out lines or curve segments connected to each other at branch points. In particular, the OBC spectrum is real if the PBC loop successfully shrinks into a segment on the real line. Our electrostatics approach below shows that this can be achieved/engineered no matter how complicated the PBC Hamiltonian is.

To connect with electrostatics (Fig.~\ref{fig:1}d), we turn to the conformal mapping $p(E)$, which is the inverse of the complex energy dispersion, the momentum $p$ and energy $E$ both regarded as complex variables. The analog of $p(E)$ in electrostatics is the quantity $\phi(z)$, where $z=x+iy$ represents real space, and curves of constant $U=\text{Re}(\phi)$ and $V=\text{Im}(\phi)$ represent field lines and equipotentials respectively. Corresponding to them are curves of constant $\text{Re}(p)=k$ and $\text{Im}(p)=\kappa$ respectively (Table~\ref{table1}). Figure~\ref{fig:1} compares the profile of the inverse skin depth $\kappa(E)$ with its analog, the electrical potential $V(z)$. \Rev{Intuitively, we make this correspondence since any conformal map, by definition, must preserve the orthogonal nature of the field lines and equipotentials. Likewise, unique isolines in the skin depth $\text{Im}(p)=\kappa$ remain orthogonal to lines of constant $\text{Re}(p)=k$ under adiabatic OBC-PBC transformation in steps of constant $\kappa$ \cite{PhysRevB.102.085151}. It should be noted that this electrostatics analogy relating the non-Hermitian skin effect problem to electrostatics is unrelated to other electrostatic analogies concerning the spectra of random non-Hermitian matrices\cite{feinberg1997non,sommers1988spectrum,haake1992statistics}, where the Green's function is found to have a $1/r$ dependence resembling the Coulomb electrostatic potential. In these works, the eigenvalue correlations are found to resemble that of a Coulomb gas, while in our analogy, it is the complex spectral flow that is found to contain the conformal structure of an equipotential vs. electric field lines pair.}
	
In addition, it is quite straightforward to associate grounded conductors, \textit{i.e.} contours of constant $V=\text{Im}(\phi)=0$ with PBC eigenenergy loops $E(k)$, $k\in[0,2\pi)$ (solid closed loops in Fig.~\ref{fig:1}c,d). This is because $\text{Im}(p)=\kappa=0$ for PBC Bloch eigenstates, due to their translation invariance. However, the OBC eigenenergies $\bar E (k)=E(k+i\kappa(k))$ do not in general correspond to any equipotential, unless $\kappa(k)$ happens to be constant. Instead, we identify the OBC spectra with regions of nonzero electric charge, since they are precisely where solution surfaces $\kappa(E)$ intersect (crests within the loops in Fig.~\ref{fig:1}c,d), such that the corresponding potential surfaces $V(z)$ become piecewise continuous. Specifically, nonzero electrical charge density $\varepsilon_0\nabla^2_zV$ is inherited from the nonzero $\nabla^2_E\kappa\neq 0$ at these crests.

		\begin{table}
		\centering
		\begin{tabular}{|p{0.25cm}|p{4cm}|p{4.2cm}|}\hline
\multirow{13}{*}{\rotatebox[origin=c]{90}{Analogues}} & \textbf{Non-Hermitian  lattice }  & \textbf{ Electrostatic system } \\ \hline

		&	Complex energy $E$  & Complex position $z=x+iy$ \\ \cline{2-3}

		&		  		\blue{\makecell{ Inverse skin depth/ \\ imaginary flux $\kappa(E)$ }}   & \blue{Electrical potential $V(z)$} \\ \cline{2-3}

				&  	\blue{Density of states $\rho_\epsilon$} & \blue{Electrostatic charge density $\sigma_\epsilon$} \\ \cline{2-3}

				 &		\makecell{ OBC to PBC spectral\\ interpolation curve} & Electric field lines of const. $U(z)$ \\ \cline{2-3}

				& 		PBC spectrum $E(k)$  & Grounded conductor $\epsilon^\text{PBC}(U)$ \\ \cline{2-3}

		 &	OBC spectrum $ E(k+i\kappa)$  & Path of nonzero charge density \\ \hline
						
		\end{tabular}
		\caption{Correspondence between a non-Hermitian lattice and its electrostatic analogy. Of primary importance are the quantities in blue, which completely define their respective systems. Analogous to the electrical potential $V$ is the skin depth $\kappa^{-1}$ of accumulated states arising from the unbalanced non-Hermitian hoppings. Whether the OBC spectrum $\bar E(k)=E(k+i\kappa)$ can be real depends on whether the Poisson equation on $V$ is consistent with nonzero charge density $\sigma_\epsilon$ only along $y=0$, \textit{i.e.} nonvanishing spectral density of states (DOS) $\rho_\epsilon$ along $\text{Im}(E)=0$. Here $\epsilon$ refers to a generic path on the complex energy (spatial) plane $E$ ($z$), emphasizing that $\sigma_\epsilon$ and $\rho_\epsilon$ are in principle proportional along any curve, not just the real line.		
}\label{table1}
	\end{table}	

	What makes this electrostatics analogy particularly useful is that the density of states (DOS) in the non-Hermitian system corresponds to the charge density in the electrostatic problem. As such, an electrostatic solution (of the Poisson equation) can be used as a proxy for solving the equivalent non-Hermitian problem, whose numerical diagonalization may be undesirably sensitive to noise.	To understand this duality between charge density and spectral density, note that the DOS along an arbitrary curve $\mathcal{\epsilon}$ 
	in the complex $E$ plane of a lattice with $L$ sites is given by
	\begin{equation}
		\rho_{\epsilon}=\frac{L}{2\pi}|\hat \epsilon\cdot\nabla_{E}(k)|=\frac{L}{2\pi}|\hat \epsilon\times \nabla_{E}(\kappa(k))|,
		\label{rho}
	\end{equation}
	since there are $L$ eigenstates along a loop $E(p)$ parametrized by $\text{Re}(p)=k\in [0,2\pi)$, with fixed $\text{Im}(p)$ chosen such that the loop intersects $\mathcal{\epsilon}$. Here $E$ and and the unit tangent vector $\hat \epsilon=\epsilon/|\epsilon|$ are treated as 2D vectors. The second equality was obtained via the Cauchy-Riemann relations. According to our electrostatics analogy, $\kappa=\text{Im}(p)$ correspond to the electrical potential $V$, and $\nabla_{E}(\kappa)$ thus corresponds to the negative electric field strength $-\mathcal{E}=\nabla V$. 
Along a path $\epsilon$ such that there is a discontinuity in the field strength, the induced charge density can be shown via a Gaussian pillbox to be 
	\begin{equation}
		\sigma_\epsilon= \mp 2\varepsilon_0|\hat\epsilon \times\nabla V|,
		\label{sigma}
	\end{equation}
	where $\varepsilon_0$ is the permittivity of the medium, and the sign of $\mp$ dependent on the direction of  the discontinuity step. Evidently, then, the charge and corresponding spectral densities are proportional, $\sigma_\epsilon\propto \rho_\epsilon$, with the proportionality constant fixed by the constraint that the total number of states in the entire spectrum is $L$.

Some comments are in order regarding the discreteness of the charges. So far, in formulating the electrostatics analogy, we have tacitly assumed large $L$, such that the DOS and hence their corresponding charge density tends towards a continuum. However, there exist scenarios of non-Bloch band collapse~\cite{alvarez2018non,longhi2020non}, where there exists couplings towards only one uncompensated direction, and the OBC spectrum shrinks into one of more isolated $\bar E$ points, with divergent $\kappa$ and hence $V$. Such scenarios exactly correspond to electrostatic problems with isolated point charges, which include extremely well-understood textbook examples. As elaborated in {Supplementary  Section~\ref{methodofimages}}, classic approaches such as the method of Images and superposition of point charges allow for the elegant design of Hamiltonians exhibiting non-Bloch band collapse onto an arbitrary variety of point charge configurations.

\subsection{Designing Hamiltonians with desired spectra via electrostatics}

We now elaborate on how our electrostatics analogy can help us engineer realistic Hamiltonians that possess desired OBC eigenenergies, particularly real eigenenergies. Conventionally, as indicated in Fig.~\ref{fig:1}b, given a lattice Hamiltonian $H(k)$ with known properties, \textit{i.e.} topological properties, one simply computes its OBC or PBC spectrum by numerical diagonalization under the respective boundary conditions. However, even though it is easy to write down a Hamiltonian $H(k)$ that gives rise to a particular PBC spectrum $E(k)$, it is much more nontrivial to design a $H(k)$ that gives a desired OBC spectrum $\bar E(k)=E(k+i\kappa(k))$, since $\kappa(k)$ usually takes a very complicated form. Yet, it is the OBC spectrum that holds the key to achieving real spectra beyond the symmetry constraint paradigm.

Our electrostatics analogy is tailor-fit to solve this inverse problem of finding $H(k)$ with desired OBC spectral properties. Based on Table~\ref{table1} and surrounding discussions, the key input specifications for the desired OBC states are exactly the crucial data for constructing the electrostatic analog. Namely, they are
\begin{enumerate}[label=(\roman*)]
\item the locus of the desired OBC spectrum in the complex $E$ plane, which specifies the locations of the analog electrostatic charges,
\item the desired spatial profile (skin depth $\kappa$) of its eigenstates, which specifies the electrical potentials at their analog charges, and
\item the shape of the desired PBC spectrum.
\end{enumerate}
As illustrated in Fig.~\ref{fig:1}, the OBC eigenvalues/charges manifest as ridges, and (i) and (ii) respectively give the positions and heights of these ridges. To uniquely define the electrostatic system, we additionally specify the boundary equipotentials that enclose these ridges, which correspond to (iii) the shape of the desired PBC spectrum.

\begin{figure*}
		\includegraphics[width=0.99\linewidth]{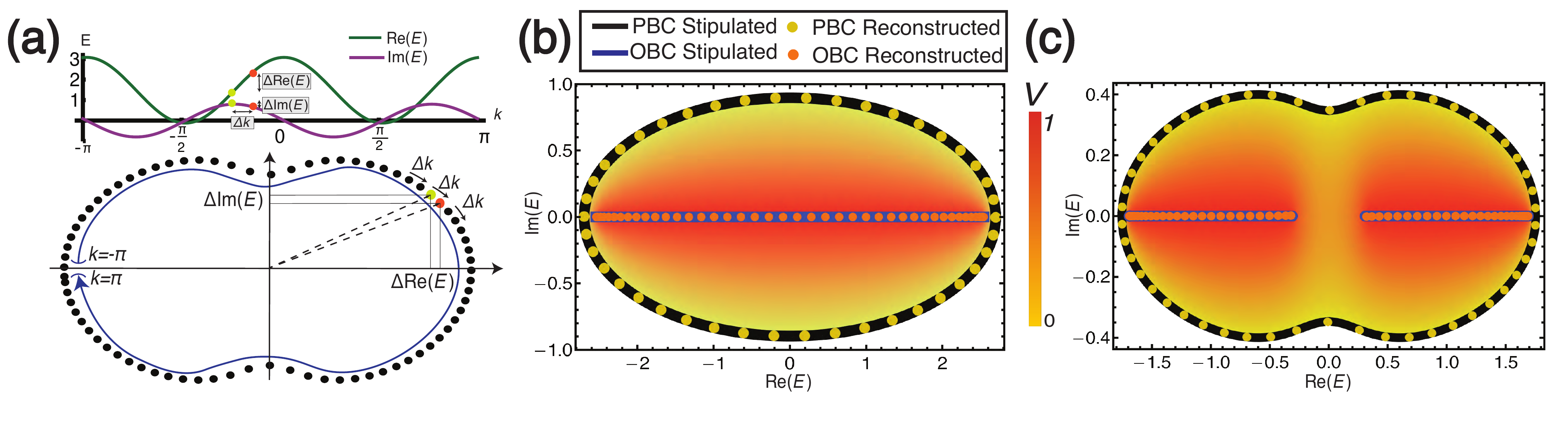}
		\caption{
\textbf{Exact agreement between the stipulated and reconstructed spectra in our design approach, for the two warm-up models. } 
\textbf{(a)} The spectrum $E(k)$ of the output Hamiltonian is reconstructed by setting energy intervals $\Delta E$ between equal momentum spacings $\Delta k$ to be inversely proportional to the density of states $\rho_\epsilon$, which is obtained from the induced charge density $\sigma_\epsilon$ in the equivalent electrostatic problem.
\textbf{(b)} Warm-up example I with stipulated OBC spectrum on the real line segment $|x|\leq 2\sqrt{t_+t_-}$ with constant eigenstate decay profile $\kappa^{-1}$, corresponding to a constant electrical potential $V$. Its elliptical PBC spectral locus corresponds to a grounded conductor, giving rise to induced charges that enable the reconstruction of the full Hamiltonian $H_\text{HN}(k)$. As a check, its PBC and OBC spectra fall exactly on the initially stipulated loci. Parameters used are $t_+ = 0.9$, $t_- = 1.8$ and $n=42$ lattice sites.   \textbf{(c)} Warm-up example II with two stipulated real OBC line segments, PBC locus of the form of Eq.~\ref{SSHeq} and constant $\kappa$. The corresponding electrical potential induces charges that allow the reconstruction of an $E_\text{SSH}(k)$ dispersion, corresponding to the 2-component $H_\text{SSH}(k)$ from Eq.~\ref{HSSH} with parameters $\gamma=0.4$, $t=0.8$ and $n=50$ sites. Both the PBC and OBC spectra of $H_\text{SSH}$ display perfect agreement with their initially stipulated loci. Note that we have normalized the potential to $V=1$ along the real line segments in both \textbf{b} and \textbf{c}.
} 
\label{fig:2}
\end{figure*}

Combined together, these data are sufficient for determining the electrostatic potential $V(z)$ everywhere, and ultimately recovering the parent $H(k)$ that produces them. This can be achieved through the following workflow, whose conceptual and implementation details are elaborated in the {Supplementary Section III}. First, we obtain $V(z)$ everywhere by solving the Poisson equation with respect to Dirichlet boundary conditions from stipulated input data (i) to (iii). Next, by applying Gauss's law on $\mathcal{E}=-\nabla V$, we obtain the induced charges on the boundary conductor, which corresponds to the PBC spectral locus. Due to the proportionality between electrostatic charge $\sigma$ and its corresponding spectral density $\rho$ (Eqs.~\ref{rho} and \ref{sigma}), we hence obtain the DOS along the PBC curve stipulated by input (iii). As illustrated in Fig.~\ref{fig:2}a, this determines the $k$-point spacing along the PBC curve, leading to a full reconstruction of $E(k)$ and hence $H(k)$ as further described in the {Supplementary Section III}. With a known candidate $H(k)$, the real-space couplings can be obtained via Fourier transformation.

For more insight into the motivation behind our approach, it is instructive to contrast generic non-Hermitian scenarios with Hermitian scenarios. In the latter, the PBC and OBC spectra are both real, and largely overlap except possibly at isolated points. Hence their potential surfaces in Fig.~\ref{fig:1}c,d would have been compressed onto the real $E$ line. This forces inputs (i) and (iii) to the same line segment on the real line. Furthermore, Hermitian eigenstates are Bloch states with $\kappa=0$, so input (ii) would have been moot, with the electrostatic setup reducing to a trivial equipotential line, instead of a potential surface in the 2D plane. At a deeper level, our electrostatics approach reinforces that even if two non-Hermitian lattice systems possess the same OBC spectra, they can still differ in two fundamentally independent aspects: their (ii) eigenstate skin depths $\kappa$ and (iii) PBC spectra.



Before presenting new Hamiltonians designed with our approach, we shall first provide two pedagogical illustrations using familiar non-Hermitian models---the Hatano-Nelson model~\cite{hatano1996localization} and the non-Hermitian SSH model~\cite{PhysRevLett.121.086803,PhysRevLett.116.133903}. Both models are already known to possess real OBC spectra due to unbalanced directed gain/loss, not symmetry protection, since the same models under PBCs do not have real spectra. However, for the purpose of illustration, we shall assume no \textit{a priori} knowledge of these models, and derive their Hamiltonians based on the input data (i) to (iii) discussed earlier: to recap, (i) range of the desired real OBC spectra, (ii) desired skin depths $\kappa^{-1}$ of their corresponding OBC states, and (iii) complex paths traced out by their PBC spectra, which uniquely specifies the possible output models.

\subsubsection{Warm-up I: Hatano-Nelson model}	

In this simplest first example, we show how the HN model (Eq.~\ref{HN}) can be recovered just from stipulating its real OBC spectrum, uniform $\kappa$ and elliptical PBC spectral locus. From the previous subsection, its OBC eigenenergies lie along the real segment $\bar E_\text{HN}\in [-2\sqrt{t_+t_-},2\sqrt{t_+t_-}]$, and the PBC eigenenergies $E_\text{HN}$ are distributed along an ellipse with semi-major/minor axes $t_+\pm t_-$ in the complex plane (Fig.~\ref{fig:2}b). Note that we have \textit{not} parametrized both the OBC and PBC spectra, because the DOS and hence functional forms of these spectra are \textit{a priori} unknown. We also stipulate that all OBC eigenstates have similar spatial decay profiles, \textit{i.e.} uniform $\kappa$ in anticipation of recovering the HN model; different distributions of $\kappa(E)$ can lead to very different parent Hamiltonians, albeit with identical OBC and PBC spectral loci.

In Fig.~\ref{fig:2}b, these stipulated OBC and PBC spectra are mapped onto an electrostatic system with a real line segment $x\in [-2\sqrt{t_+t_-},2\sqrt{t_+t_-}]$ (blue) of elevated uniform potential $V=\log\sqrt{t_-/t_+}$ enclosed by an elliptical equipotential $V=0$ (black) with semi-major/minor axes $t_+\pm t_-$. The Laplace equation on this boundary-value problem possesses an analytic solution $\smash{V(z) = -\log|(z\pm \sqrt{z^2-4t_+t_-})/2t_-|}$, the $\pm$ sign depending on $\text{sgn}[x(t_--t_+)]$. This gives precisely the functional form of $V=\text{Im}(p)$ in the (complexified) energy dispersion $E_\text{HN}(p)=z$, establishing that energy $E=E_\text{HN}$ indeed corresponds to the position $z$ in this analogy, and that $H_\text{HN}$ is indeed the parent Hamiltonian that yields these stipulated specifications on the spectrum and $\kappa$.

\subsubsection{Warm-up II: non-Hermitian SSH model}	

We next show how a more complicated 2-component parent Hamiltonian can be recovered via our approach without the benefit of a known analytic solution. As a warm-up demonstration, the stipulated spectral loci are based on the known non-Hermitian SSH model~\cite{PhysRevLett.121.086803,PhysRevLett.116.133903} $H_\text{SSH}$, even though the model itself is assumed to be \textit{a priori} unknown, and meant to be reconstructed. In this case, the real spectrum $\bar E_\text{SSH}$ is specified to lie within the two real segments $1-T<|\text{Re}(E)|<1+T$, corresponding to the segments of constant elevated potential $V$ in the $z$ plane of the electrostatic analog (Fig.~\ref{fig:2}c). They are enclosed by a grounded ($V=0$) outer boundary corresponding to the PBC locus of $E_\text{SSH}$, which we stipulate (with the benefit of hindsight, for illustration purposes) as 
\begin{equation}
\gamma^2(x^2-y^2+\gamma^2-1-t^2)^2=4t^2(\gamma^2-x^2y^2),
\label{SSHeq}
\end{equation}
$\gamma$ and $t=\sqrt{T^2+\gamma^2}$ parameters of the $H_\text{SSH}$ to be found. Spatial coordinates $(x,y)$ correspond to the spectral locus $(\text{Re}(E_\text{SSH}),\text{Im}(E_\text{SSH}))$. We emphasize that specifying the locus of $E_\text{SSH}$ does not imply knowledge of either the PBC spectrum $E_\text{SSH}(k)$ or its Hamiltonian $H_\text{SSH}(k)$, since the parametrization with $k$ is still unknown.

For a model with 2 or more components, obtaining a physically realistic Hamiltonian $H(k)$ from the dispersion $E(k)$ via $\text{Det}[H(k)-E(k)\mathbb{I}]=0$ presents an additional avenue of subtlety. As there are many possible forms of $H(k)$ which all yield the same eigenenergy $E(k)$, it is crucial to choose a correct ansatz that allows all components of $H(k)$ to possesses rapidly decaying Fourier coefficients, such that the solution corresponds to a local Hamiltonian. In general, the Fourier coefficients $f_x$ of a function $f(k)$ decay like  $f_x\sim x^{-(1+\beta)}e^{-\eta x}$, where $\eta$ is the distance between the real line and the nearest complex singularity of $f(k)$ and $\beta$, the order of that singularity, which is either negative or fractional. That is, $f(k+i\eta)\approx(k-k_0)^\beta+f_0$, with $k_0$ and $f_0$ being constants~\cite{he2001exponential,PhysRevB.91.085119}. In particular, local Hamiltonians should possess $\eta>0$, which means $H(k)$ should not contain divergences or branch points at $k\in \mathbb{R}$.


In this case, the reconstructed PBC dispersion can be fitted to the square-root expression $E_\text{SSH}(k) =\pm \sqrt{1+2t\cos k+T^2 +2i\gamma\sin k}$, which will result in unphysical long-ranged hoppings if we use a 1-component ansatz Hamiltonian. But since $E_\text{SSH}^2(k)$ does not contain branch cuts, a prudent choice would be a 2-component ansatz Hamiltonian with only off-diagonal elements (see {Supplementary Section III}), which can be shown to be
\begin{equation}
H_\text{SSH}(k)= (t+\cos k)\sigma_x+(\sin k + i\gamma)\sigma_y.
\label{HSSH}
\end{equation}
Shown in Fig.~\ref{fig:2}c is the excellent match between the OBC/PBC spectra of the reconstructed $H_\text{SSH}$ (orange/yellow), compared with the initially stipulated elevated potential interval along the real line (blue) and the boundary equipotential (black). This is one of the very few models where a constant $\kappa$ deformation, \textit{i.e.} potential $V$, can recover a real OBC spectrum---specifically, $k\rightarrow p = k+i \kappa$ where $\kappa=\log[(t+\gamma)/(t-\gamma)] / 2$, such that $\bar E_\text{SSH}(k) = E_\text{SSH}(p)=\pm \sqrt{1+2T\cos k +T^2 }$ is real when $|t|>|\gamma|$. In the following, we shall explore how, by varying the PBC spectral locus and the $\kappa(E)$ profile for OBC states, we can engineer a much greater variety of hitherto undiscovered models with real spectra.




	\begin{figure*}
		\includegraphics[width=\linewidth]{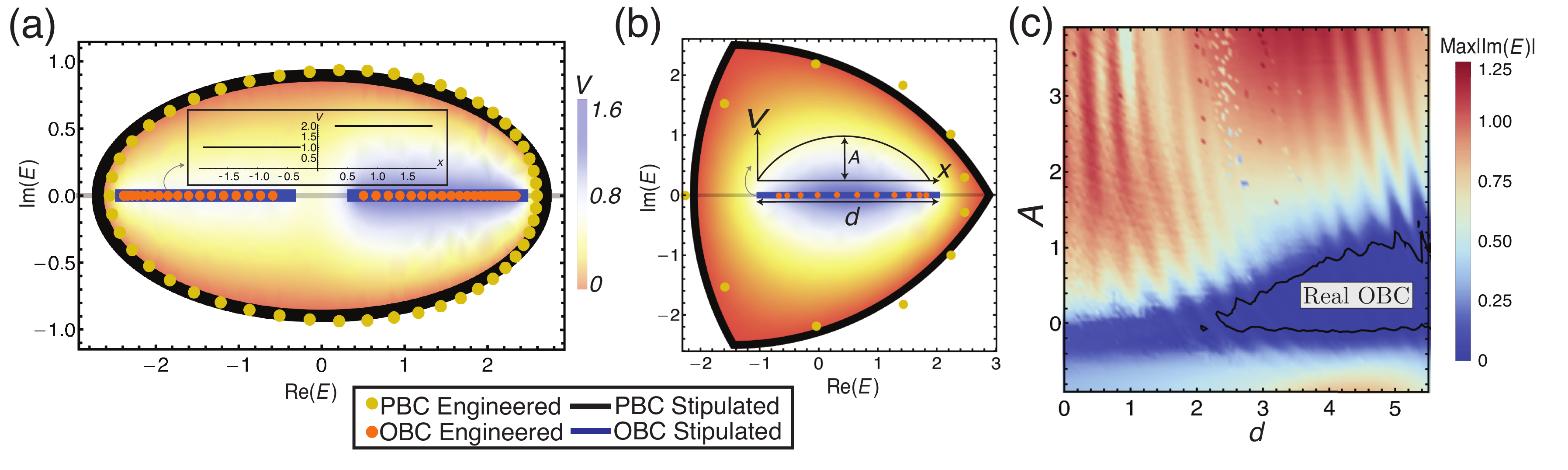}
		\caption{\textbf{Illustrative examples demonstrating a search for real-spectra parent Hamiltonians in nontrivial settings without special symmetries. } \textbf{(a)} PBC ellipse with two real OBC segments with unequal skin depths: the stipulated real OBC segments $\pm E \in[0.3,2.4]$ are of asymmetrically different skin depths $\kappa^{-1}=1$ and $1/2$, corresponding to potentials $V=1$ and $1.6$. Together with grounded conductors defined by the stipulated PBC spectral locus $(\Re(E)/2.7)^2+(\Im(E)/0.9)^2=1$, they define an electrostatic problem, yielding a parent Hamiltonian in the form of Eq.~\ref{E2ansatz}. \textbf{(b)} PBC Reuleaux triangle with real OBC segment of length $d$ and position offset $x_0=-1$, with inhomogeneous skin depths corresponding to a concave potential $V(x)$ of amplitude $A$ (Eq.~\ref{Vx}). Engineered (dotted) and stipulated (solid) spectra agree well, even though the engineered Hamiltonian is single-component with only up to next-nearest neighbor hoppings. \textbf{(c)} Values of $\text{Max}|\text{Im}(E)|$ for the OBC spectrum of our engineered Hamiltonian in the parameter space of $A$ and $d$. There exists a large parameter region (blue) with almost real OBC, which is not expected given the very different symmetries of the Reuleaux triangle and the $V(x)$ profile. Tolerance for reality is $\text{Max}|\text{Im}(E)| \leq 0.02$, and $20$ sites are used. 
\\
}
		\label{fig:3}
	\end{figure*}

\subsubsection{Real OBC states with non-constant skin depths $\kappa^{-1}$}	

We now turn to the first nontrivial demonstration of our approach, where we design a Hamiltonian stipulated to possess an OBC spectrum occupying two real line segments, each with a different value of $\kappa$. Consider the spectra shown in Fig.~\ref{fig:3}a, which naively looks like a simple variant of the Hatano-Nelson or SSH model, with specified OBC real line segments lying symmetrically about the origin, surrounded by an also symmetric elliptical PBC spectral locus. However, what is nontrivially asymmetric is the unequal inverse skin depth $\kappa$ (corresponding to $V$) of the eigenstates lying on each line segment, which cannot be fulfilled via any simple extension of these models. As a property with no Hermitian analog, $\kappa$ is entrenched in the complex analytic properties of the Hamiltonian, and it is nontrivial to tune it without also modifying the spectrum. This is when our electrostatics analogy becomes valuable: corresponding to $\kappa$ is the electrical potential $V$, and the task is reduced to that of solving the Laplace equation, subject to these $V=1$ and $V=1.6$ segments and the $V=0$ outer boundary. 

Upon solving this electrostatic problem and reconstructing the corresponding dispersion $E(k)$, we find that $E(k)$ has a slowly decaying train of Fourier components, and cannot result from a single-component local Hamiltonian. As discussed in the previous subsection, this is generically expected, and a resolving strategy is to consider an engineered Hamiltonian with 2 or more components (bands). To accommodate the asymmetry, we generalize the previous 2-component ansatz to:
\begin{equation}
H(k)=H_z(k)\sigma_z+T(k)\mathbb{I}+\mathcal{F}(k)\sigma_++\sigma_-,
\label{E2ansatz}
\end{equation}
where $\sigma_\pm=(\sigma_x\pm i \sigma_y)/2$, which allows the reconstructed energy dispersion $E(k)$ to be fitted to either branch of the expression
\begin{equation}
E(k) = T(k) \pm \sqrt{\mathcal{F}(k)+H_z(k)^2},
\end{equation}
where $T(k)$, $H_z(k)$ and $\mathcal{F}(k)$ each contain no more than a few Fourier components. The details of the fitting and solution branch selection are given in the {Supplementary Section IV}. 

Showcased in Fig.~\ref{fig:3}a is the very good agreement between the PBC and the OBC spectra of the engineered Hamiltonian, compared to their stipulated counterparts. For this example, up to next-nearest neighbor hoppings were kept in $H$ (see {Supplementary Section VI}); even better agreement and reality of spectra can obtained with further hoppings. Notice the unequal eigenenergy spacings between the left and right halves, which result ``naturally'' as induced charges from the asymmetric electrical potential (pale orange-purple). This asymmetric example also illustrates the distinction between the PBC dispersion $E(k)$ and the PBC spectral locus, which is the ellipse that is initially specified. Even though the latter is known \textit{a priori}, the actual dispersion and density of states, which is skewed to the right as shown in Fig.~\ref{fig:3}a, can only be known after the potential profile $V(x,y)$, \textit{i.e.} $\kappa(E)$, has been solved for. Graphically, it is also evident that steeper potential gradients lead to stronger induced charges and hence DOS on the right half. In the {Supplementary Section VII}, it is shown that good agreement was obtained not just for the stipulated and engineered spectra, but also for the $\kappa(\text{Re}(E))$ profile, which corresponds to the $V(x)$ potential.

\subsection{Real spectra without any symmetry}
\label{sec:results/real-no-symmetry}

Even though our electrostatics approach can in principle generate a parent Hamiltonian corresponding to any desired spectral loci and skin depth profiles, the usefulness of the reconstructed Hamiltonian hinges on its experimental feasibility. As such, it is oftentimes preferable to restrict the reconstructed Hamiltonians to a small number of relatively local real-space hoppings. With further hoppings truncated, the compromise is that the reconstructed spectra may no longer be exactly real, even though the stipulated desired spectra lie along the real line. Even then, our approach allows for a systematic exploration of the region in parameter space where a class of local hopping models possesses almost real spectra, unaided by any specific symmetry.

As a demonstration, consider the rather demanding scenario where the desired PBC spectrum is stipulated along a so-called Reuleaux triangle (Fig.~\ref{fig:3}b), which is parametrically defined by the expression $E_\text{Reuleaux}(t)=\sqrt{3} e^{\frac{1}{6} i \left(2 \pi  \left\lfloor \frac{3 t}{2 \pi }\right\rfloor +3 t+\pi \right)}-e^{\frac{1}{3} i \pi  \left(2 \left\lfloor \frac{3 t}{2 \pi }\right\rfloor +1\right)}$, $t\in [0,2\pi)$. Such a shape has 3-fold rotational symmetry about the complex origin, which generically do not encourage real OBC spectra. Meanwhile, the stipulated OBC spectral locus is set to be a segment of length $d$ along the real line $x\in (-1,2)$, although it remains to be seen how faithfully the reconstructed OBC spectrum can reproduce this. As an additional complication, $\kappa(\text{Re}(E))$ (which corresponds to potential $V(x)$) is set to 
\begin{equation}
V(x)=1+A\sin\left[\frac{\pi(x-x_0)}{d-x_0}\right]
\label{Vx}
\end{equation}
with offset $x_0 = -1$ for definiteness, and $A,d$ parameters to be tuned. We proceed as in the previous examples, and is able to obtain local and simple Hamiltonians that produce almost exactly real OBC spectra while respecting the stipulated inputs of this problem. For instance, the spectra showcased in Fig.~\ref{fig:3}b arose from the engineered Hamiltonian
\begin{equation}
H=\sum_j 0.613c^\dagger_jc_j-2.18c^\dagger_jc_{j+1}-0.193c^\dagger_{j+1}c_j-0.5c^\dagger_jc_{j+2}
	\label{eqn:9}
\end{equation}
whose real spectrum possess eigenstate localization profiles closely obeying $V(x)$, despite hoppings with no apparent symmetry whatsoever that even suggests of the possibility of a real spectrum. \Rev{Containing only up to next-nearest neighbor hoppings, it is simple enough to realize in photonic systems given recent developments in the capability to control long range  hoppings \cite{bouganne2020anomalous,song2019breakup,el2019dawn}. Mechanical and ultracold atomic systems also provide lee ways into implementing such systems \cite{PhysRevLett.125.118001,qian2022ultracold}. But perhaps the most versatile means is mechanical or electrical systems where hoppings of arbitrary order can be generated with appropriate circuit configurations\cite{Helbig2020,gu2016holographic,lenggenhager2021electric}.}


The propensity of real spectra can be increased by optimizing parameters $A$ and $d$. Shown in Fig.~\ref{fig:3}c is the value of OBC $\text{Max}|\text{Im}(E)|$ in the parameter space of $A$ and $d$, with hoppings across up to 5 sites. In general, although the reconstructed OBC spectra are not always perfectly real, they are still mostly real with ''branches'' pointing away from the real line (See Fig.~\ref{fig:S2}). Notably, there exists a large parameter region (purple) where the OBC spectrum is almost perfectly real, with $\text{Im}(E)$ at least two orders of magnitude smaller than $\text{Re}(E)$. Examples of almost-real spectra at different points in the parameter space can be found in the {Supplementary Section VIII}.

It is remarkable that this Hamiltonian can possess any real OBC spectrum at all, given that it not only lacks suitable conventional symmetry such as hermiticity or PT-symmetry, but also lacks any other ostensible symmetry that can encourage PBC$\rightarrow$OBC spectral flow towards the real line.

	\begin{figure}
		\includegraphics[width=0.99\linewidth]{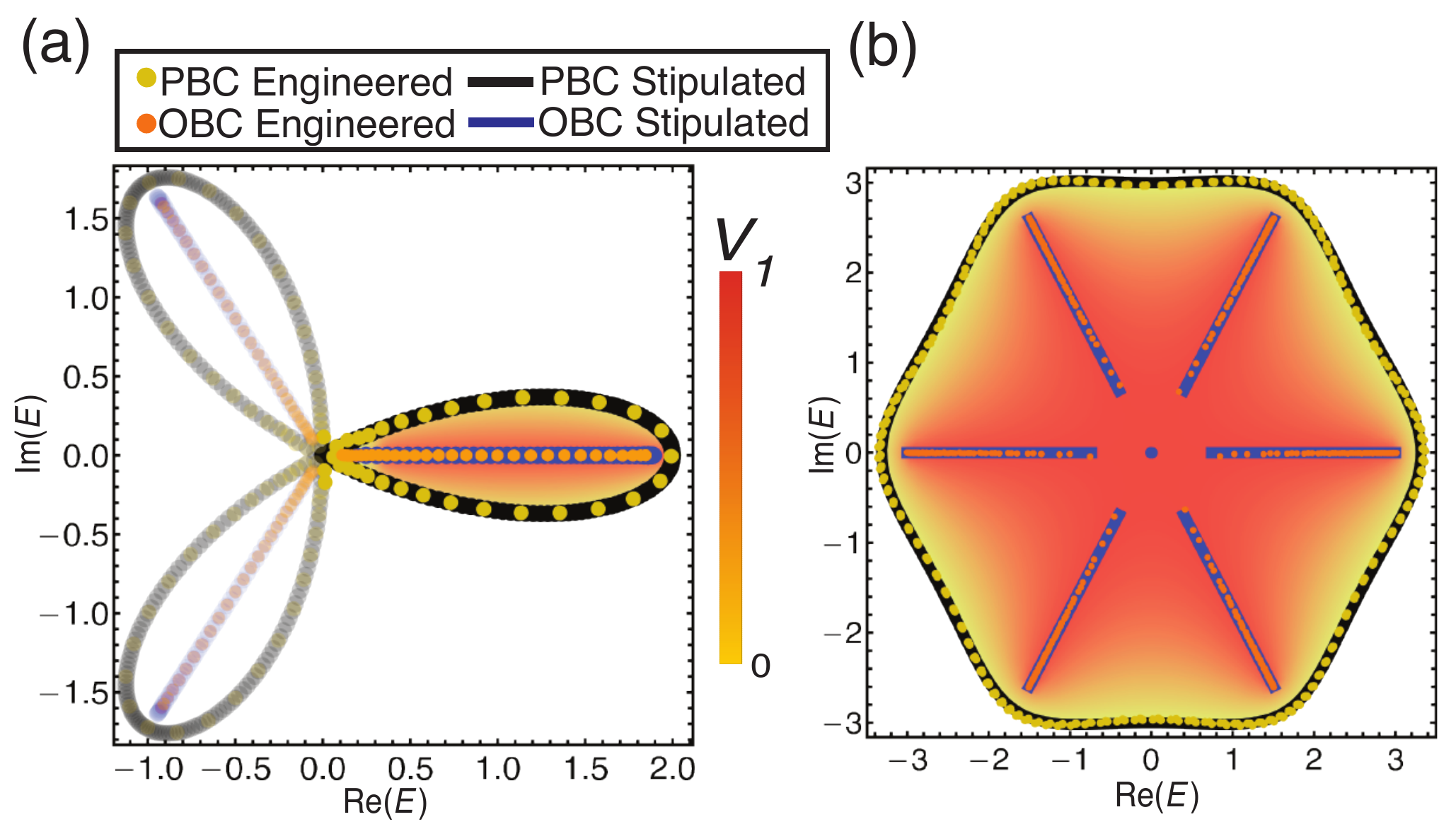}
		\caption{\textbf{Designing beyond real spectra. } \textbf{(a)} Almost perfect agreement between the stipulated and engineered spectra in the right lobe was achieved, with only up to next-nearest neighbor hoppings, by discarding solutions from the two other extraneous lobes. The stipulated PBC spectrum is given by the parametrization Eq.~\ref{Et3lobe}. \textbf{(b)} Excellent reconstruction can also be achieved by enlarging the number of components of the ansatz Hamiltonian and its solution branches. Here, the 3-component ansatz $H_\text{6-fold}$ (Eq.~\ref{E3ansatz}) with $\Delta =0$ allows for an excellent local fitting of $E(k)^3$ with minimal Fourier components. Stipulated OBC and PBC spectra are given by Eqs.~\ref{4b1} and \ref{4b2}, with $r_\text{min}=0.9835$, $r_\text{max}=2.9532$. We used $30$ and $120$ lattice sites for \textbf{(a)} and \textbf{(b)} respectively, the large number in the latter necessary for demonstrating excellent convergence in all the branches.}
		\label{fig:4}
			\end{figure}
				
\subsection{Design of generic spectra from electrostatics}
	
While most of this work is focused on the engineering of new classes of model Hamiltonians with real spectra, our approach is equally useful for obtaining those with desired non-real spectra of any shape, since the electrostatics analogy remains valid for PBC or OBC spectra, \textit{i.e.} boundary potentials of any shape. We hereby describe two additional applications, where only certain branches of the OBC spectra need to be real.

\subsubsection{Hamiltonians with real spectral branch}
	
When desired states can be selectively filled, it suffices that the spectrum possesses a real branch which can be occupied, such that the reality of other branches become immaterial. Relaxing the requirement for the \textit{entire} spectrum to be real greatly enlarges the space of candidate Hamiltonians. Shown in Fig.~\ref{fig:4}a, for instance, is a simple example where the stipulated real OBC branch $x \in (0,1.9)$ lies within a lobe-shaped stipulated PBC loop parametrized by 
\begin{equation}
E(t)=e^{it}+e^{-2it},\qquad |t|<\pi/3.
\label{Et3lobe}
\end{equation}
 Considered in isolation, the reconstructed Hamiltonian would have required many non-local terms due to the cusp in the PBC loop. However, if this PBC lobe and real OBC branch were part of the full 3-fold symmetric spectrum (faded in Fig.~\ref{fig:4}a), the solution to $E(k)$ would be simply given by the expression of $E(t)$ above.


\subsubsection{Multi-component Hamiltonians}
With this example in mind, we can repeat our methodology for more sophisticated multi-component models. Here, the entire stipulated OBC spectra is not restricted to the real line. For definiteness, suppose we desire to have \textit{both} the OBC and PBC spectra exhibit approximately straight line segments, such that we can get various real spectra sectors upon suitable energy translation or rotation. In Fig.~\ref{fig:4}b, the stipulated OBC spectrum is given 6 straight line segments 
\begin{equation} r_\text{min}\leq|z|\leq r_\text{max},\qquad \arg(z)=\frac{n\pi}{3},n\in\mathbb{Z},\label{4b1}\end{equation}
related by $\pi/3$ rotations, together with a zero mode, while the stipulated PBC spectrum is parametrized by
\begin{equation}E^2(t)=ae^{-it}+e^{2it},\label{4b2}\end{equation} 
which is a sixfold rotationally symmetric figure with approximately straight sides when $a=10$. It should be noted that this $t$ parametrization will not correspond to the expected PBC dispersion relation $E(k)$ of the engineered Hamiltonian, since only the locus of points needs to be matched.

From the geometric symmetry of the spectra, we chose a 3-component ansatz for the engineered Hamiltonian of the form
\begin{equation}H_\text{6-fold}(k)=\begin{pmatrix}\Delta & f(k) & 0 \\ 0 & \Delta & 1 \\ 1 & 0 & \Delta\end{pmatrix},\label{E3ansatz}\end{equation}
with dispersion $E(k)=\Delta + e^{2\pi i n/3}\sqrt[3]{f(k)}$, $n=0,1,2$, such that $f(k)$ can be found from the truncated Fourier transform of the reconstructed $(E(k)-\Delta)^3$ obtained from our induced electrostatic charge solutions. For Fig.~\ref{fig:4}b, we have set the energy shift $\Delta$ to be zero; we could have set $\Delta=3i$ had we wanted to have a segment of approximately real PBC spectrum. As shown, keeping only hoppings within the same unit cell and across 3 unit cells in $f(k)$ (see {Supplementary Section VI}), the reconstructed spectra agree with the stipulated spectra almost perfectly. This would not have been possible had we chosen another ansatz that involves some root of $f(k)$ other than $\sqrt[3]{f(k)}$.


	\subsection{Discussion and conclusion}
	In a handful of memorable pages on electrostatic analogs and the unity of nature \cite{Feynman}, Richard Feynman pointed out how many different phenomena in physics can all be explained using the equations of electrostatics. These include the heat flow between plates held at different temperatures, the vibrations of a drumhead, the diffusion of neutrons and the flow of a fluid past a sphere. Here we unravelled a new electrostatics analogy from non-Hermitian condensed matter physics and inverse quantum engineering. Specifically, we rigorously prove that the seemingly hard problem to synthesize non-Hermitian spectra in tight-binding lattices with modified bulk-boundary correspondence can be mapped onto a simple electrostatic problem. The working principle behind this analogy is the intimate relation between the conformal structure of electrostatics and the complex spectral flow in non-Hermitian systems. 
Our work opens up a new paradigm for engineering non-Hermitian spectra, particularly real spectra, in various settings, such as cold atoms~\cite{bouganne2020anomalous,song2019breakup}, 
photonics~\cite{ElGanainy2018,Longhi2017}, metamaterials~\cite{PhysRevLett.124.073603,pickup2020synthetic}, mechanical and acoustic systems~\cite{PhysRevLett.125.118001}. While real spectra are important for state stability in the majority of experiments, we point out that non-real spectra present further possibilities in terms of topological sophistication~\cite{PhysRevB.102.085151,zhang2022real}, and are just as physically relevant in the form of the Laplacian spectra of steady-state networks such as electrical circuits~\cite{ningyuan2015time,Lee2018,Helbig2020,lenggenhager2021electric,shang2022experimental}. 
	
\Rev{Our approach will impact the design of non-Hermitian sensors\cite{doi:10.1126/science.aar7709} in optics and electronics. As finite-sized devices, they are OBC systems, and the requisite sensing properties would specify the non-Hermitian skin state profiles, which correspond to the potential profile of the analog electrostatic charges.  Such sensors may find applications in measuring glucose concentrations\cite{PhysRevApplied.11.044049} and wireless sensors\cite{Zeng_2021}.} 

Besides its versatility in discovering new classes of stable non-Hermitian Hamiltonians, our electrostatics analogy also allows numerical access to non-Hermitian lattices of far larger system sizes. Till now, the numerical computation of OBC skin spectra have been limited by rapidly compounding floating point errors, which limit accurate diagonalization results to systems smaller than $\mathcal{O}(10^2)$ sites at standard machine precision. Analytic results only exists for a small subset of systems where the generalized Brillouin zone is not excessively complicated. However, our electrostatics approach trades this non-Hermitian diagonalization problem with a boundary-value partial differential equation problem, whose numerical solution do not suffer from non-Hermitian sensitivity at all, and can be extended even towards the thermodynamic limit.

\subsection{Author Contributions}
CHL developed the electrostatics analogy and initiated the project. CHL and RY led the project. JWT performed the spectral reconstruction. RY, JWT, TT and JMK performed the numerical computations. SL generalized the electrostatics analogy to point charges. SL, LL and CHL also took on advisory roles. All authors contributed to the writing of the manuscript.

\subsection{Acknowledgements}
This work is supported by Singapore's MOE Tier I grant WBS no. A-800022-00-00.

\bibliographystyle{ieeetr}

\bibliography{ref}	

\clearpage

\begin{center}
\textbf{\large Supplementary Sections}\end{center}
\setcounter{equation}{0}
\setcounter{figure}{0}
\setcounter{table}{0}
\setcounter{section}{0}
\renewcommand{\theequation}{S\arabic{equation}}
\renewcommand{\thefigure}{S\arabic{figure}}
\renewcommand{\thetable}{S~\Roman{table}}
\renewcommand{\cite}[1]{\citep{#1}}

	\section{Obtaining real OBC spectra through the dynamic cancellation of gain and loss}
	
In the Hatano-Nelson model discussed in the main text, we have seen the simplest instance of real spectra originating not from symmetry constraints, but from the cancellation of unbalanced gain and loss as the state is pumped against a boundary. To rigorously understand this mechanism, we review the physics of state pumping due to unbalanced gain/loss, formally known as the non-Hermitian skin effect (NHSE)~\cite{PhysRevLett.121.086803,PhysRevB.99.201103}. Consider a 1D system with $L$ sites governed by a Hamiltonian $H=\sum_k H(k)c^\dagger_kc_k$. Its PBC eigenenergies $E$ are simply given by the momentum eigenstate solutions of the characteristic polynomial (energy dispersion equation) $\text{Det}[H(k)-E\,\mathbb{I}]=0$, where $k=2n\pi/L$, $n\in\mathbb{Z}$. Under OBCs and other scenarios with broken translation symmetry, however, eigensolutions must vanish at both boundaries, and must thus consist of a linear combination of at least two of such ``momentum'' eigenstates - which agree with the true momentum eigenstates only between the open boundaries, but vanish elsewhere. Yet for NHSE systems with unbalanced couplings, such degenerate eigensolutions for the same OBC eigenenergy $\bar E$ are spatially decaying, and are thus characterized by momenta $p(k)=k+i\kappa(k)$ with nonvanishing imaginary parts which we shall call $\text{Im}(p(k))=\kappa(k)$. For their linear combination to exist in the thermodynamic limit, without one solution dominating the other, we thus require degeneracy also in the exponent of $\psi_k(x)\sim e^{-\kappa(k) x}$, \textit{i.e.} degeneracy of $\kappa(k)$. Such OBC solutions are exponentially decaying away from the boundaries with localization lengths $\kappa^{-1}(k)$, and are thus known as skin states. Specifically, we are concerned with engineering arbitrarily complicated $H(k)$ such that the OBC eigenenergies $\bar E$ of $\bar H(k)=H(k+i\kappa(k))$ are real.

	\section{Overview of spectral reconstruction workflow} 
		\label{appendix:A}

	In this section, we present a more detailed run-through of our workflow in engineering the parent Hamiltonians of desired spectra. The inputs and outputs of our engineering scheme are summarized in Table~\ref{table2}. As introduced in the main text, the inputs represent the desired features of the Hamiltonian to be engineered, namely the complete specification of the OBC spectrum $\bar E$ and eigenstate decay rate (skin depth $\kappa^{-1}$), as well as the shape (locus) of the PBC spectrum $E\in\epsilon^\text{PBC}$ in the complex energy plane. Note that specifying the PBC spectral locus does not entail specifying the PBC energy dispersion $E(k)\in\epsilon^\text{PBC}$---in the latter, the parametric dependence on the momentum $k$ also contains information on the density of states $|\nabla_k E|^{-1}$, and allows the construction of possible parent Hamiltonian $H(k)$ consistent with the dispersion. Likewise, the OBC Hamiltonian is also not completed specified with the input data, since its DOS is not given. The inverse skin depth $\kappa(E)$ associated with each eigenenergy $E$ on the OBC spectral locus $E\in\epsilon^\text{OBC}$ represents the complexification of the Bloch momentum, and only exists in non-Hermitian systems with broken translation symmetry, \textit{i.e.} OBCs.
	
Our engineering workflow is designed to leverage readily available tools in solving the Poisson equation in electrostatics. Based on the three desired aspects of the non-Hermitian lattice system that are \textit{a priori} specified, we can unambiguously specify an analogous electrostatic system. As summarized in Table~\ref{table1} of the main text, the PBC spectral locus $\epsilon^\text{PBC}$ in the complex $E$ plane corresponds to the shape of the grounded equipotential, \textit{i.e.} conductor in the electrostatic analog. The OBC spectral data does not correspond to any equipotential in general, since OBC eigenstates are characterized by the inverse skin depth $\kappa^{-1}(E)$, which correspond to non-constant potential $V(z)$ on $\epsilon^\text{OBC}$. In essence, our input data of the desired non-Hermitian model features completely specifies the corresponding electrostatic problem by specifying all the boundary potentials (at $\epsilon^\text{PBC}$ and $\epsilon^\text{OBC}$).

This boundary value problem can be directly treated via a numerical Laplace equation solver. In the equivalent electrostatic problem, the effects of the charges are encapsulated in the boundary potentials $V(z)$. Solving it, we recover the (output) potential $V(z)$ everywhere. While the non-Hermitian analog to $V(z)$ is $\kappa(E)$, which is not a very useful quantity by itself, what is useful is the induced charge on the conductors $\sigma_{\epsilon}$,  $\epsilon=\epsilon^\text{PBC}$ due to the output $V(z)$. In particular, the spectral density of states $\rho$ is proportional to electrostatic charge $\sigma$. Together with the shape of $\epsilon^{\text{PBC}}$ on the complex $E$ plane, we can reconstruct the complex energy dispersion $E(k)$ and hence its parent Hamiltonian through techniques elaborated below. The profile of $V(z)$ similarly allows us to recover the DOS of OBC states, but that cannot be directly used for reconstructing the parent Hamiltonian.

\begin{table}
		\centering
		\begin{tabular}{|l|l|l|}\hline
		&	\textbf{ Nonherm. lattice }  & \textbf{ Electrostatic system } \\ \hline
	\textbf{Input}	&	OBC spectral loci $\epsilon^\text{OBC}$ & Location of given charges \\ \hline
	\textbf{Input}	&	\makecell{OBC skin depth  \\ $\kappa^{-1}(E)$ on $\epsilon^\text{OBC}$ } & Potential at given charges \\ \hline
	\blue{\textbf{Input}}	&	\blue{PBC spectral loci $\epsilon^\text{PBC}$}  & \blue{Grounded conductor shape} \\ \hline
		\blue{\textbf{Output}} & \blue{DOS of PBC states}  & \blue{Ind. charge on conductor} \\ \hline
  	\textbf{Output} & DOS of OBC states  & Values of given charges \\ \hline
	\textbf{Output}	&	$\kappa(E)$ profile everywhere & Potential $V(z)$ everywhere \\ \hline
\end{tabular}
		\caption{Summary of input and output quantities of our spectral engineering workflow, together with their electrostatic analogs. Blue quantities are directly involved in recovering the engineered parent Hamiltonian. The inputs specified are desired features of engineered model, which include a complete specification of the OBC states and target PBC eigenenergies. The latter, combined with their corresponding density of states (DOS) which is output from our workflow, completely specifies the PBC and hence parent Hamiltonian. As a bonus, our solution recovers the potential $V$ for the electrostatic analog, as well as its sources/sinks (charges) everywhere.
}\label{table2}
	\end{table}
	
	\begin{figure*}
		\includegraphics[width=0.99\linewidth]{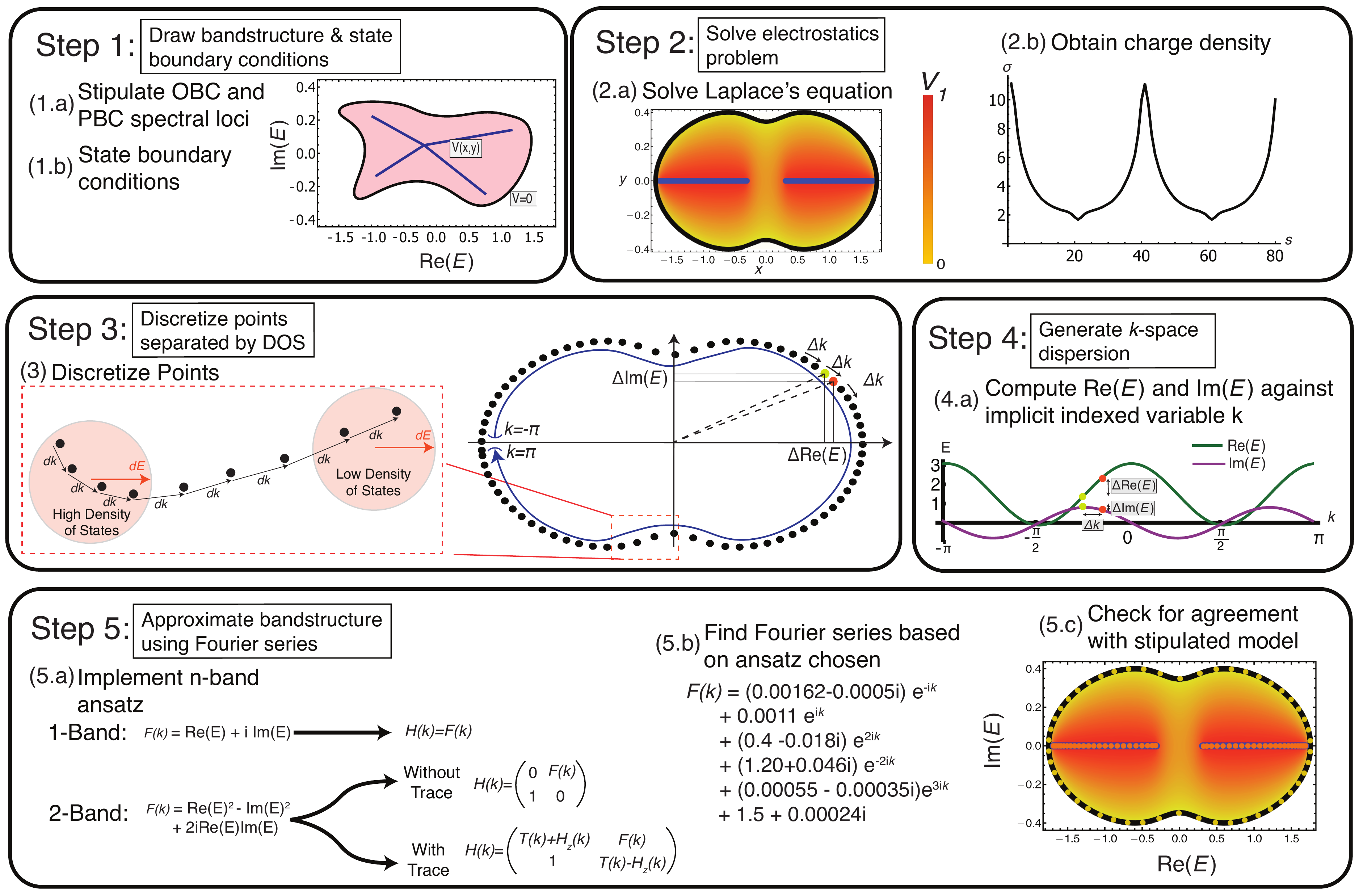}
		\caption{Scheme of work for reconstructing the (engineered) parent Hamiltonian given stipulated PBC/OBC spectral loci and skin depths, which define the input boundary conditions. Two illustrative ansatz Hamiltonians are shown; further possibilities are discussed in the text.}
		\label{fig:S1}
	\end{figure*}
	
\section{Workflow details}

Below, we do an explicit walk-through of our workflow, starting from specifying the problem to obtaining the final reconstructed Hamiltonian and its spectra. For illustration purposes, we shall base this demonstration on the non-Hermitian SSH model (warm-up example II in the main text), with stipulated input data generated from parameters $\gamma=0.4$ and $t=0.8$.

\subsection{Defining the input by stipulating the desired spectra \textit{i.e.} band structure/boundary conditions}
	

In the thermodynamic limit $L \rightarrow \infty$, the PBC and OBC spectra trace out loci given respectively by a continuous PBC loop and an OBC graph. While an actual lattice possess only a finite $L$ number of points, 
a continuum approximation is crucial in defining the electrostatic problem, such that the  charge density $\sigma_\epsilon$, and by extension the density of states (DOS) $\rho_\epsilon$ is properly defined everywhere. 

By definition, the PBC loop corresponds to $\kappa=V=0$, and is thus represented by a grounded conductor that defines the outer boundary of the electrostatic problem (outer loop in Step 1 of Fig.~\ref{fig:S1}). The OBC spectral locus lies within the PBC loop, and generically corresponds to non-constant potential $V(x,y)$ (schematic blue branching lines in Step 1), as stipulated by the OBC inverse skin depths, even though it is constant for the non-Hermitian SSH model. This OBC potential forms another set of boundary potentials.

	

	\subsection{Solving the Electrostatic Problem}
	
	After the electrostatic problem is completely specified via its boundary potentials, the Laplace equation \begin{equation}\laplacian{V(x,y)}=0,\end{equation} is solved within the region between the boundaries (PBC and OBC spectral loci), via a finite-element method scheme. This is a unique Dirichelet problem with open curves inside a closed loop (OBC loci)~\cite{krutitskii2000dirichlet}, and there is no charge away from the OBC and PBC loci, since all charges only reside along the OBC loci, where the potential is allowed to be discontinuous. Importantly, as a fixed-size problem treated by efficient finite-element solvers, our approach is applicable for arbitrarily large systems, even in the thermodynamic limit, unlike the explicit computation of non-Hermitian spectra through diagonalization, which can run into convergence issues due to the NHSE even with only $\mathcal{O}(10^2)$ sites.
	
After obtaining the potential $V(x,y)$ everywhere along the interior of the PBC loop, one may then calculate the electric field $\vec{\mathcal{E}}=-\nabla V$. By Gauss's law, the charge density $\sigma_\epsilon$ along the PBC boundary is given by
		\begin{equation}
		\sigma_\epsilon = \vec{\mathcal{E}}\cdot \hat{n},
	\end{equation}
where $\hat{n}$ is the vector normal to the PBC loop (grounded conductor), pointing inwards toward the OBC loci. For the purpose of obtaining the DOS $\rho_\epsilon\propto \sigma_\epsilon$ along the PBC loop, we compute and plot $\sigma_\epsilon$ along the arc length $s$ of the PBC loop. In 2b of Fig.~\ref{fig:S1}, we observe two peaks and troughs for $\sigma_\epsilon$ over a period of $L=80$ sites, consistent with the 2-fold rotational symmetry of the setup.
	
In our implementation, this step is integrated with the following steps 3 and 4 into an automated routine. The engineering of Hamiltonians given desired specifications is thus made simple, without the need for manual intervention. Details on the code are available upon request.
	
	
	\subsection{Discretization of Points as specified by density of states/charges}

The next step in obtaining the dispersion $E(k)$ is to infer the rate of change of $E$ on $k$ along the PBC loop, from density of states (DOS) data. In our electrostatics analogy, this DOS $\rho_\epsilon$ is conveniently proportional to the previously computed charge density $\sigma_\epsilon$. 

We partition the PBC loop with $L$ points, such that the separation between the points are inversely proportional to the DOS $\rho_\epsilon=\Delta k/\Delta E$. In $k$-space,  each point is thus separated by a fixed momentum interval $\Delta k=L/2\pi$, and the set of all points therefore gives the PBC spectrum evaluated $E(k)$ at $L$ equally spaced $k$ points. In step 3 of Fig.~\ref{fig:S1}, regions of low/high DOS, \textit{i.e.} charge correspond to points that are further/closer together. 
It should also be noted that it is irrelevant where the $k=0$ reference point is placed, and simply starting at any point in the loop and picking a consistent traversal direction is sufficient. 
	
	\subsection{Generating the momentum dispersion}

Through simple interpolation, the functional forms of $\textrm{Re}(E(k))$ and $\textrm{Im}(E(k))$ are computed, from the numerical data at the $L$ equally-spaced momentum points. If necessary, the higher powers of the reconstructed $E$ can be similarly computed.
	
\subsection{Reconstructing the Hamiltonian through Fourier series approximations}
\label{subsubsection: FS Approx}
In this final step, we engineer a reasonably local Hamiltonian that yields the spectrum $E(k)$ reconstructed from the DOS data. First, an ansatz Hamiltonian $H(k)$ needs to be selected, which can be any $n\times n$ component (band) matrix. For practicality, however, we would like $n$ to be as small as possible, subject to the constraint that the components of $H(k)$ contain a small number of Fourier components. Higher-order Fourier components correspond to distant couplings, thus allowing too many of them ruins the locality of the Hamiltonian, and is experimentally undesirable.

	\begin{figure*}
		\includegraphics[width=0.99\linewidth]{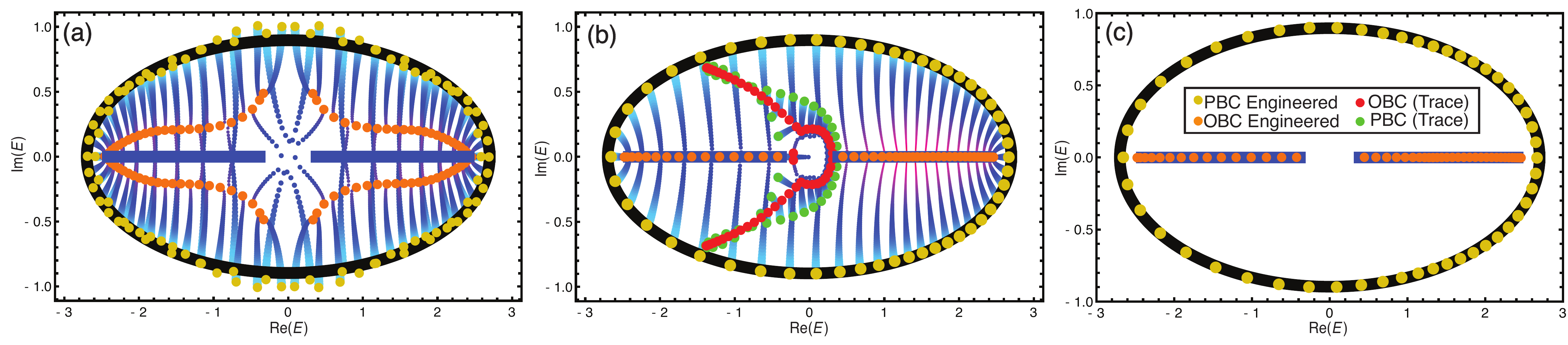}
		\caption{ Example use of a trace term in drastically improving convergence to the stipulated real spectrum. \textbf{(a)} With a traceless ansatz $H(k)=\mathcal{F}(k)\sigma_{+}+\sigma_{-}$ with $\mathcal{F}(k)$ given by Eq.~\ref{E3ansatz}, there is poor agreement between stipulated (blue) and engineered (orange) OBC spectra, particularly at the discontinuity centered at the origin. \textbf{(b)} Upon implementation of an optimized ansatz with trace term as detailed in Eqns.~\ref{E2ansatz} and \ref{E3ansatz}, the spectral flow from the PBC to purely real OBC line becomes what is desired, with the two solution branches morphing into a pair of purely real spectra segments (orange), accompanied by 
unwanted eigenenergy bands shown in green and red for PBCs and OBCs respectively.  \textbf{(c)} After discarding the unwanted eigenenergies, we reach an almost exact agreement between the stipulated and reconstructed real spectra.} 
		\label{fig:S4}
	\end{figure*}	

	\begin{figure*}
		\includegraphics[width=0.9\linewidth]{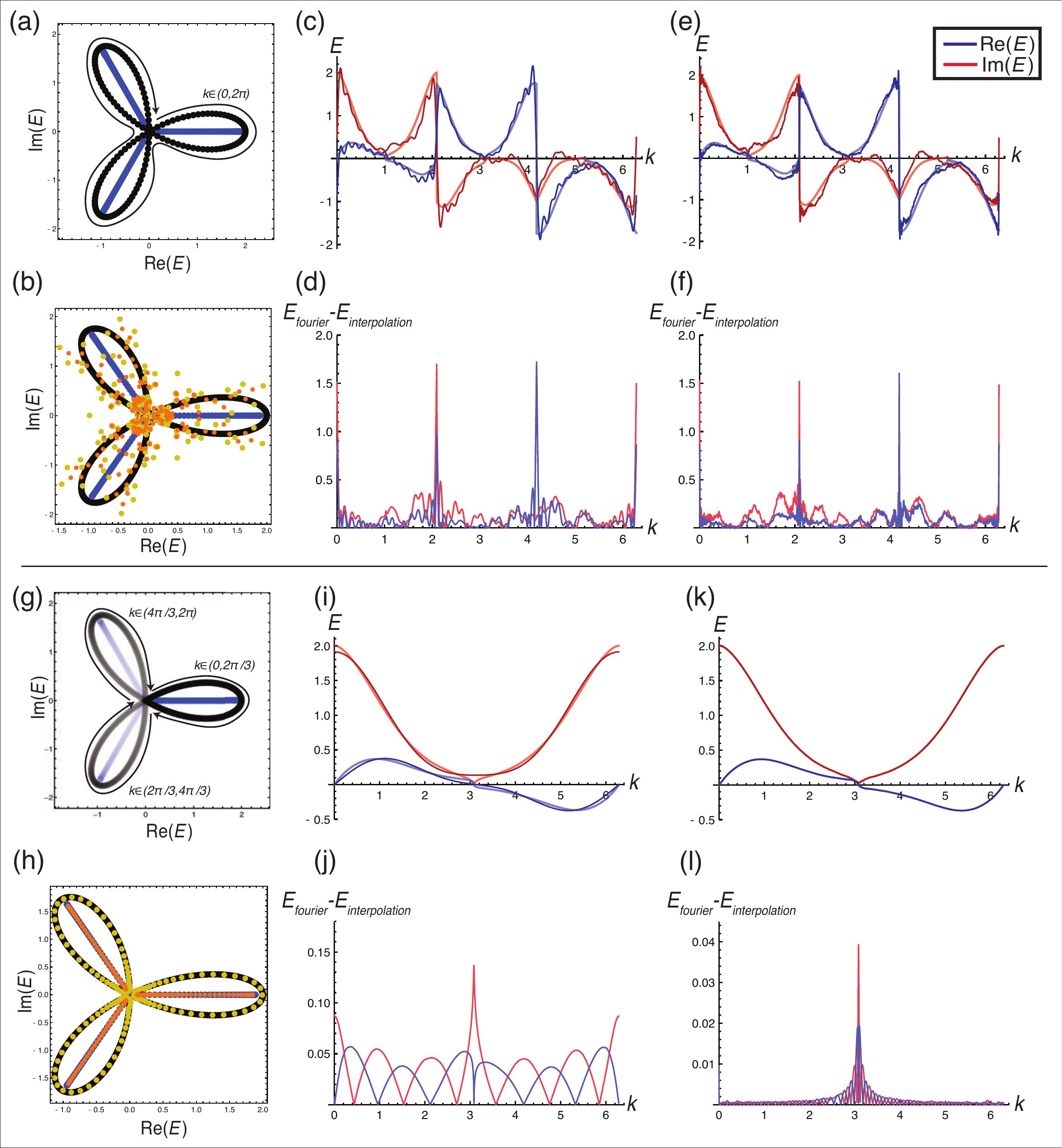}
		\caption{\textbf{(a)-(f)} Convergence without segmenting bandstructure symmetrically as compared to \textbf{(g)-(l)} where rotational symmetry was used. \textbf{a} does not segment the 3-fold rotationally symmetric bandstructure and uses a continuous domain of $k\in(0,2\pi)$. \textbf{b} demonstrates the poor convergence when implemented where yellow points represent PBC and orange points represent the OBC complex bandstructure. A stark difference is observed in \textbf{g} and \textbf{h}. \textbf{c} and \textbf{e} show the use of 40 and 400 Fourier terms respectively while \textbf{d} and \textbf{f} show the difference between the Fourier series approximation and the interpolation. At the sharp jump, the Gibbs's phenomenon can be observed where the difference converges to a finite value instead of 0. This is contrasted by \textbf{i} and \textbf{k} where only 2 out of 20 Fourier terms are used are used and \textbf{j} and \textbf{l} show that the difference in the approximation is 1-2 orders smaller than the preceding case.}
		\label{fig:S2}
	\end{figure*}

While a few scenarios may already be satisfactorily solved with a 1-component model, usually at least 2 components are needed. This is to prevent unnecessary branching of the OBC locus caused by slow power-series convergence under a 1-component model. When the reconstructed dispersion $E(k)$ cannot be expressed as the sum of a few Fourier terms, better convergence can often be achieved by taking higher powers of $E(k)$ or its truncations. 
As such, by using a multi-component ansatz Hamiltonian, one arrives at matrix elements that are functionally dependent on the dispersion in ways that could result in far enhanced convergence. Two examples of 2-component ansatzes are given in step 5a of Fig.~\ref{fig:S1}. Here, we discuss the simpler one with $H(k)=((0,\mathcal{F}(k)),(1,0))$, where there is only one undetermined matrix element $\mathcal{F}(k)$ to be fitted, and $\mathcal{F}(k)=E(k)^2$. This demonstrative ansatz has been chosen for simplicity - in well-known examples like the SSH model, both off-diagonal elements are present, which usually makes for easier physical realization. For the particular example featured in Fig.~\ref{fig:S1}, this ansatz with a single functional degree of freedom $\mathcal{F}$ proves sufficient, since the numerically fitted Fourier series (step 5b) contains only a few powers of $e^{ik}$ with non-negligible coefficients, and corresponds to the sum of a few local couplings. 

After obtaining the engineered (output) Hamiltonian, one may verify (step 5c) that its PBC and OBC spectra indeed agrees well with the originally stipulated (input) spectra. One can then be confident that numerical and truncation errors accumulated in the solution are acceptably small. In Fig.~\ref{fig:S1}, we observe that, for our illustrative system, the agreement is virtually perfect. 

\section{Approaches to Improving Convergence}
\subsection{Exploiting additional degrees of freedom in constructing Hamiltonians}


In many cases, the dispersion $E(k)$ does not correspond to a rapidly decaying Fourier series, thereby precluding a local single-component parent Hamiltonian. An arbitrarily high number of Fourier coefficients may be required to achieve the desired accuracy.

However, a multi-component Hamiltonian with local matrix elements provide the requisite degrees of freedom that satisfies the same dispersion. 
As previously discussed, a simplest 2-component extension goes by the ansatz $H(k)=\mathcal{F}(k)\sigma_{+}+\sigma_{-}$, which works when $E(k)^2=\mathcal{F}(k)$ instead of $E(k)$ is local. However, this form is still too restrictive, applicable to only a special subset of models.

By introducing a trace term $T(k)=\Tr(H(k))\neq 0$, we can expand the possible functional forms of characteristic polynomials $f(E, e^{ik})$ that can be achieved i.e. 
to that of the more general form $E^2-T(k)E+\det(H(k))=0$, with two independent functional degrees of freedom $T(k)$ and $\det(H(k))$ in deciding the form of the Hamiltonian. With added algebraic complexity, this can analogously be extended to Hamiltonians of higher bands.

While a multi-component Hamiltonian gives rise to multiple energy bands, only one band can correspond to the desired i.e. stipulated spectrum. In many experimental realizations, the other bands can be left unoccupied and thus ignored. In implementing $T(k)$, one finds that typically only terms up to the next-nearest hoppings are needed to converge closely to the desired Hamiltonian. In Fig. \ref{fig:S4}, we show how this can be performed to achieve a branch of almost perfectly real OBC spectrum, with the Hamiltonian ansatz being given by $H(k)=H_z(k)\sigma_z+T(k)\mathbb{I}+\mathcal{F}(k)\sigma_++\sigma_-$ (Eqn. \ref{E2ansatz}) with:
\begin{align}
	T(k) =& 0.7 e^{ik}+0.4 e^{-ik} \nonumber
	\\
	\mathcal{F}(k) =& 0.08 + 0.024 e^{ik} + 0.0004 e^{-ik}\nonumber \\&- 0.08 e^{2ik} - 0.018 e^{-2ik}
	\\
	H_z(k) =& 0.362+ 1.7 e^{ik}+0.8\nonumber e^{-ik}\\&-0.227e^{2ik}-0.126 e^{-2ik}\nonumber
	\label{E3ansatz}
\end{align}



Ultimately, the choice of ansatz is also heavily influenced by practical physical considerations, such as the structure and nature of the physical lattice used to realize the engineered Hamiltonian. For instance, asymmetric couplings are easily realized in classical electrical circuits, but less straightforwardly in quantum circuits~\cite{blais2021circuit, kollar2019hyperbolic}. 

	\subsection{Exploiting n-fold rotational symmetry}
	
	In cases where the spectra exhibit $n$-fold rotational symmetry in both the PBC loop and OBC loci, one may expand the scope of candidate Hamiltonians by partitioning the energy spectrum into equally spaced intervals of $k=2\pi/n$ where $n$ is the order of symmetry by rotation, such that one interval coincides with the stipulated spectrum. We demonstrate this method with the Hamiltonian $H=z+1/z^2$ previously presented in Fig \ref{fig:4}. Visually, the PBC spectra consists of 3 equally sized "lobes" offset by a rotation $2\pi/3$. One may break each lobe into separate momentum intervals $(0,2\pi/3),(2\pi/3,4\pi/3),(4\pi/3,2\pi)$. The benefit of doing so is illustrated and elaborated in Fig.~\ref{fig:S2}. By segmenting the fourier series into piecewise separate piecewise functions, one avoids the Gibbs phenomenon---where arbitrarily large partial sums fail to converge point-wise to the target function.

\section{Non-Bloch band collapse and the method of images}
\label{methodofimages}

\begin{figure}[!h]
\centering\includegraphics[width=2.7in]{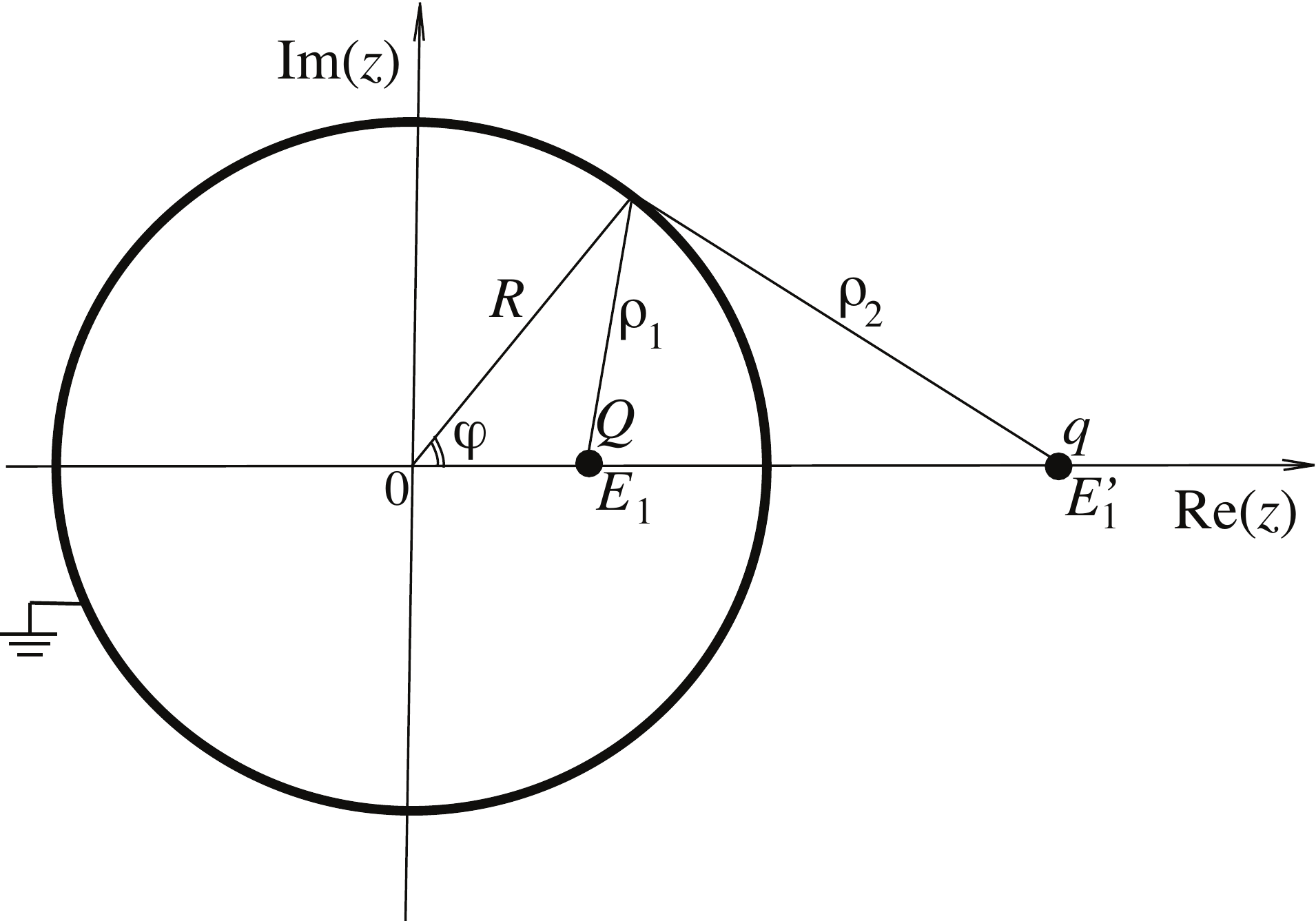}
\caption {\textbf{Method of images and the synthesis of a NH lattice displaying non-Bloch band collapse}. A point charge $Q$ is placed in the hollow region of a grounded (zero-potential) 2D conductor surface (the circle of radius $R$). The charge is placed at a distance $E_1$ from the center. The electrostatic problem is solved by considering a virtual (image) point charge $q$, in the exterior of the circle, at a distance $E_1^{\prime}$.}
\label{FigMIm}
\end{figure}

An interesting limiting case is the one where the OBC energy spectrum collapses to one point $E=E_1$ (or to a set of points) in the interior of the PBC energy spectrum, corresponding to so-called non-Bloch band collapse or flat band limit. In the electrostatics analogy,  the energy $E=E_1$ is a singularity of the potential $V$ (vanishing of the skin depth or $\kappa=V= \infty$) due to a point charge placed at the point $z=z_1=E_1$. A useful analytical tool to solve certain electrostatic problems involving point charges in the presence of conductors is the method of images, where insertion of additional virtual point charges makes the conductor surface isopotential. This method solves, for example, the problem of a point charge or an electrostatic dipole placed in front of a plane (flat) conductor or in the interior of a sphere. In the spirit of our analogy, the method of images can be harnessed to synthesize a NH Hamiltonian displaying a non-Bloch band collapse (flat band) at some target real energy $E_1$. For illustrative purposes, let us synthesize using the method of images a NH Hamiltonian $E=H(p)$, with complex momentum $p=k+i \kappa$, such that the PBC energy spectrum, i.e. $E=H(p)$ with $\kappa=0$ and $p=k$, describes the circle of radius $R$, centered at the origin of the complex energy plane, whereas the OBC energy spectrum collapses to the real energy $E=E_1$ in the interior of the circle, i.e. $-R<E_1<R$ (see Fig.~\ref{FigMIm}). In the electrostatics analogy, the complex momentum $p$ corresponds to the complex potential $\phi$, 
i.e. $p=\phi$, whereas energy $E$ corresponds to the complex spatial position $z=x+i y$, i.e. $E=z$. We are thus faced with the problem of finding a complex potential $\phi(z)=U(z)+iV(z)$ satisfying the following requirements (i) $\phi(z)$ shows a singularity at $z=E_1$; (ii) $\nabla_z^2 \phi=0$ in the interior $|z|<R$ and $ z \neq E_1$, (iii) on the circle $|z|=R$ the potential $V(z)$, i.e. imaginary part of $\phi(z)$, must vanish (grounded conductor). From the well-known form of the complex potential of a point charge in two dimensions, we can satisfy the above requirements by looking for a solution of the form
\begin{equation}
\phi(z)=i \frac{Q}{2 \pi} \log (z-E_1)+i \frac{q}{2 \pi} \log (z-E_{1}^{\prime} )+i \phi_0
\end{equation}
where $Q$ is the point charge at the singularity $z=E_1$, internal to the conductor, $q$ is the image charge, external to the conductor and placed at a distance $E_1^{\prime}$, and $\phi_0$ is a constant potential term (see Fig.~\ref{FigMIm}). Clearly, $\nabla_z^2 \phi(z)=$ for $|z|<R$ and $z \neq E_1$. The image point charge $q$, its distance $E_1^{\prime}$ from the center of the conductor and the constant potential term $\phi_0$ should be determined by imposing conditions (ii) and (iii). Clearly, on the conductor $z=R \exp(i \varphi)$ one has
\begin{equation}
V(\varphi)={\rm Im}  \{ \phi(z=R \exp(i \varphi)) \}=\frac{Q}{2 \pi} \log \rho_1+\frac{q}{2 \pi} \log \rho_2+\phi_0 \label{cazz0}
\end{equation}
where
\begin{eqnarray}
\rho_1^2 & = & R^2+E_1^2-2RE_1 \cos \varphi \label{cazz1} \\
\rho_2^2 & = & R^2+E_1^{\prime 2}-2R E_1^{\prime} \cos \varphi \label{cazz2}
\end{eqnarray}
are the squares of the distances $\rho_1=|z-E_1|$, $\rho_2=|z-E_1^{\prime}|$ of the two point charges from the point $z=R \exp(i \varphi)$ on the conductor. Eliminating $\varphi$ from Eqs.\ref{cazz1} and \ref{cazz2}, one obtains
\begin{equation}
\rho_2^2=R^2+E_1^{\prime 2}+ \frac{E_1^{\prime}}{E_1} ( \rho_1^2-R^2-E_1^2).
\end{equation}
In order for $V(\varphi)$ to be independent of $\varphi$, we set $R^2+E_1^{\prime 2}-E_1^{\prime}(R^2+E_1^2)/E_1=0$, i.e.
\begin{equation}
E_1^{\prime}=\frac{R^2}{E_1} \label{cazz3}
\end{equation}
so that $\rho_2/ \rho_1= R/E_1$, and
\begin{equation}
q=-Q. \label{cazz4}
\end{equation}
Under such conditions, from Eq.\ref{cazz0} one obtains
\begin{equation}
V(\varphi)=-\frac{Q}{2 \pi} \log  \left(\frac{R}{E_1} \right)+ \phi_0
\end{equation}
so that condition (iii) is met by letting $\phi_0=(Q/ 2 \pi) \log (R/E_1)$. The complex potential $\phi(z)$ thus reads
\begin{equation}
\phi(z)=i \frac{Q}{2 \pi} \log \left( \frac{R(z-E_1)}{E_1z-R^2} \right). \label{cazz5}
\end{equation}
The Hamiltonian $E=H(p)$ is finally obtained by solving Eq.\ref{cazz5} with respect to $z$, and letting $z=E=H(p)$, $\phi=p$. This yields
\begin{equation}
H(p)= \frac{R \exp(- 2 \pi i p /Q) -E_1}{(E_1/R) \exp(-2 \pi i p/Q)-1}.
\end{equation}
The point charge $Q$ should be chosen such that, for $p=k$ real, the exponential function $\exp( - 2 \pi i p /Q)$ is invariant for $k \rightarrow k+ 2 \pi$. A possible choice is $Q=2 \pi$, so as one finally obtains
\begin{equation}
H(p)= \frac{R \exp(-  i p) -E_1}{(E_1/R) \exp(- i p)-1}.
\end{equation}
In physical space, this Hamiltonian is realized in a lattice with hopping amplitudes $t_n$ defined by the Fourier series $H(k)=\sum_n t_n \exp(ik n)$, i.e.
\begin{equation}
t_n= \frac{1}{2 \pi} \int_{-\pi}^{\pi} dk H(k) \exp(-i kn). \label{cazz6}
\end{equation}
In the limiting case $E_1=0$, i.e. when the OBC energy spectrum collapses at the center of the PBC energy spectrum, one has $H(p)=-R \exp(-ip)$, which basically describes the Hatano-Nelson model in the limit of a unidirectional nearest-neighbor hopping $R$. For $E_1 \neq 0$, there are non-nearest hopping. To compute $t_n$, one can calculate the integral on the right hand side of Eq.\ref{cazz6} by the residue theorem. For $n \geq 0$, after letting $\xi=\exp(-ik)$ one has
\begin{equation}
t_n=-\frac{1}{2 \pi i} \oint_{|\xi|=1} d \xi \; \xi^{n-1} \frac{R \xi-E_1}{(E_1/R) \xi-1}= \left\{  
\begin{array}{cc}
E_1 & n=0 \\
0 & n \geq 1
\end{array}
\right.
\end{equation}
For $n<0$, after letting $\xi=\exp(ik)$ one obtains
\begin{eqnarray}
t_n & = & \frac{1}{2 \pi i} \oint_{|\xi|=1} d \xi \; \xi^{-n-1} \frac{R-E_1 \xi}{(E_1/R)-\xi} \nonumber \\
& = & 
\left( \frac{R}{E_1}\right)^{n+1} \left( \frac{E_1^2}{R} -R \right).
\end{eqnarray}
Interestingly, for $E_1 \neq 0$ the hopping in the lattice remains unidirectional, since $t_n=0$ for $n>0$ while $t_n \neq 0$ for $n \leq 0$, with $|t_n| $ exponentially decaying toward zero as $n \rightarrow -\infty$. The matrix Hamiltonian in physical space is thus an upper triangular matrix with $E_1$ on the main diagonal. Under OBC, the $L \times L$ matrix is defective with all eigenvalues collapsing to the value $E_1$, which is a $L$-order exceptional point, $L$ being the number of lattice sites.~\\
Finally, we mention that, owing to the superposition principle, the previous result can be readily extended to the case of an arbitrary number $N$ of point charges in the interior of the circle, placed at the complex energies (positions) $E_1$, $E_2$,..., $E_N$. The corresponding synthesized lattice shows a non-Bloch collapse under OBC to the energies $E_1$, $E_2$,..., $E_N$.  In this case, assuming again $Q= 2 \pi$, the complex potential $\phi(z)$ reads
\begin{equation}
\phi(z)=i \sum_{n=1}^N \log \left( \frac{R(z-E_n)}{E_nz-R^2} \right) 
\end{equation}
and the Hamiltonian $H=H(p)$ is implicitly defined as a root of the algebraic equation
\begin{equation}
\exp(-ip) \prod_{n=1}^N \left( E_n H(p)-R^2\right)= R^N \prod_{n=1}^N \left( H(p)-E_n \right).
\end{equation}

\section{Details of engineered Hamiltonians}
This section gives explicit expressions for the various engineered Hamiltonians designed in the main text, namely, those in Fig.~\ref{fig:3}a, \ref{fig:3}b and \ref{fig:4}b. These Hamiltonians were engineered via our electrostatics approach, based on stipulated spectral loci and inverse state localization lengths $\kappa$ that were used as inputs. 

For the OBC and PBC spectra recovered in Fig.~\ref{fig:3}a, the engineered Hamiltonian takes the following ansatz
\begin{equation}  H(k)=H_z(k)\sigma_z+T(k)\mathbb{I}+
\mathcal{F}(k)\sigma_+ +\sigma_-
\end{equation}
with trace and off-diagonal terms truncated to 
\begin{multline}
T(k) = 0.7 e^{ik}+0.4 e^{-ik}
   \\
\mathcal{F}(k) = 0.08 + 0.024 e^{ik} + 0.0004 e^{-ik} - 0.08 e^{2ik} - 0.018 e^{-2ik}.
	\\
H_z(k) = 0.362+ 1.7 e^{ik}+0.8 e^{-ik}-0.227e^{2ik}-0.126 e^{-2ik}
\label{Eqn:DisjointedOBC}
\end{multline}
Due to the degrees of freedom afforded by the presence of both $T(k)$ and $\mathcal{F}(k)$, good agreement between the stipulated and engineered spectra is achieved with hoppings of up to next-nearest neighbors only.

The engineered Hamiltonian featured in Fig.~\ref{fig:3}b takes the form
\begin{multline}
    H(k)=0.613 -2.18 e^{-i k}-0.193 e^{i k}-0.5 e^{-2 i k},
\end{multline}
and gives a good fit with the stipulated real OBC spectrum despite being a simple single-component model with up to next-nearest neighbor couplings. This example one of the several further presented below, in Fig.~\ref{fig:S4}, where different parameters for the real OBC spectral segment was stipulated.

Finally, the engineered Hamiltonian for Fig.~\ref{fig:4}b is $H_\text{6-fold}(k)=$
\begin{equation}
   \begin{pmatrix}
    0 & (5.67-0.309 i)e^{3ik}+(31.7 - 1.67 i)e^{-3ik} & 0 \\ 0 & 0 & 1 \\ 1 & 0 & 0
    \end{pmatrix}
\end{equation}
which exhibits excellent stipulated vs. engineered spectra agreement, and contains hoppings only within the same unit cell and across three unit cells.

	\begin{figure*}
		\includegraphics[width=\linewidth]{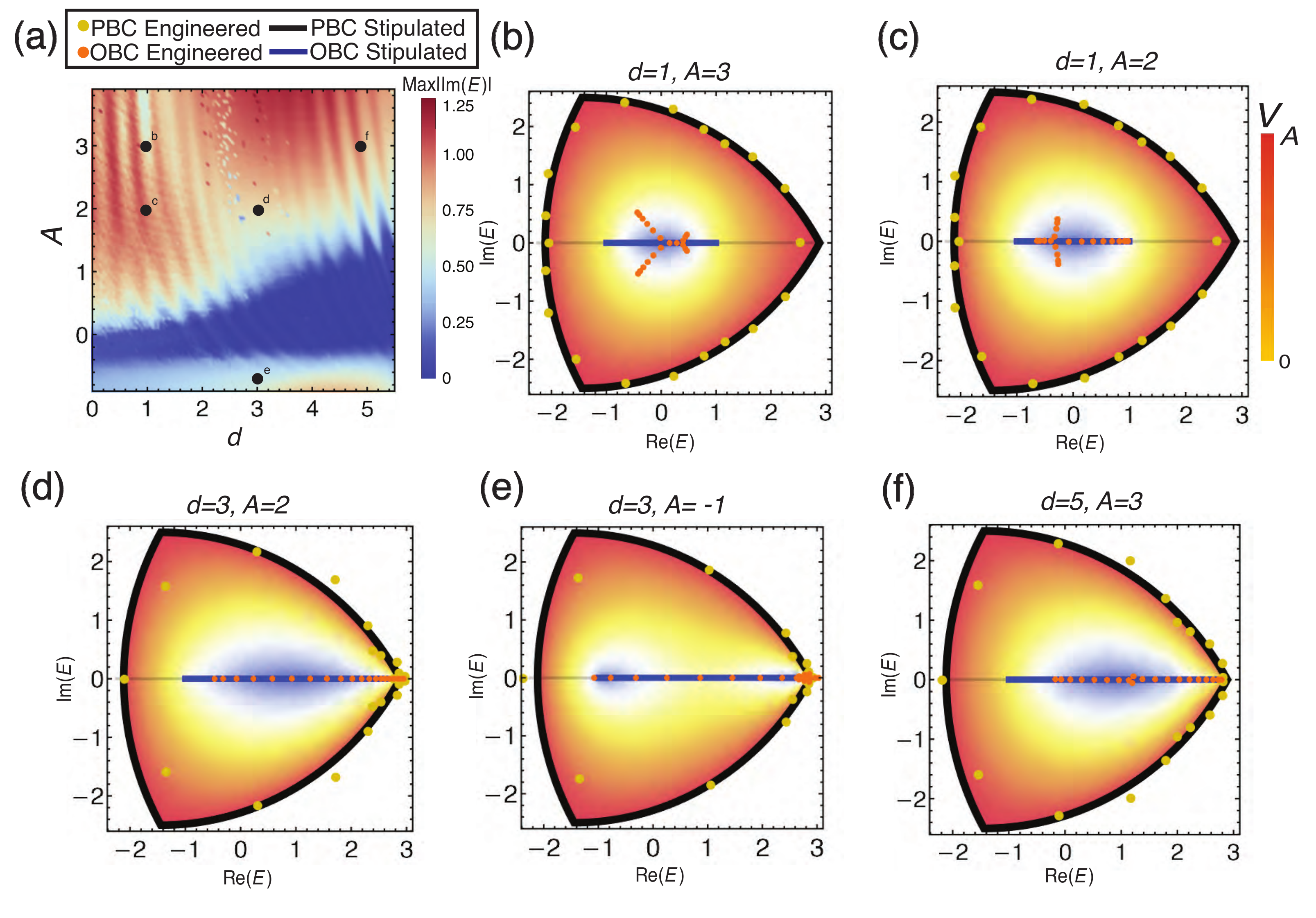}
				\caption{\textbf{a} The magnitude of $\text{Im}(E)$ in the engineered OBC spectrum in the parameter space of $A$ and $d$, as stipulated by $V(x)$ (Eq.~\ref{Vx} of the main text). The engineered spectra for various combinations of $(A,d)$, indicated by the black dots, are plotted in \textbf{b-f}. As we can see, almost perfectly real spectra are achieved in some cases.}
		\label{figS4}
	\end{figure*}

	\section{Verification of $\kappa$-profile}
	
	\begin{figure*}
		\includegraphics[width=\linewidth]{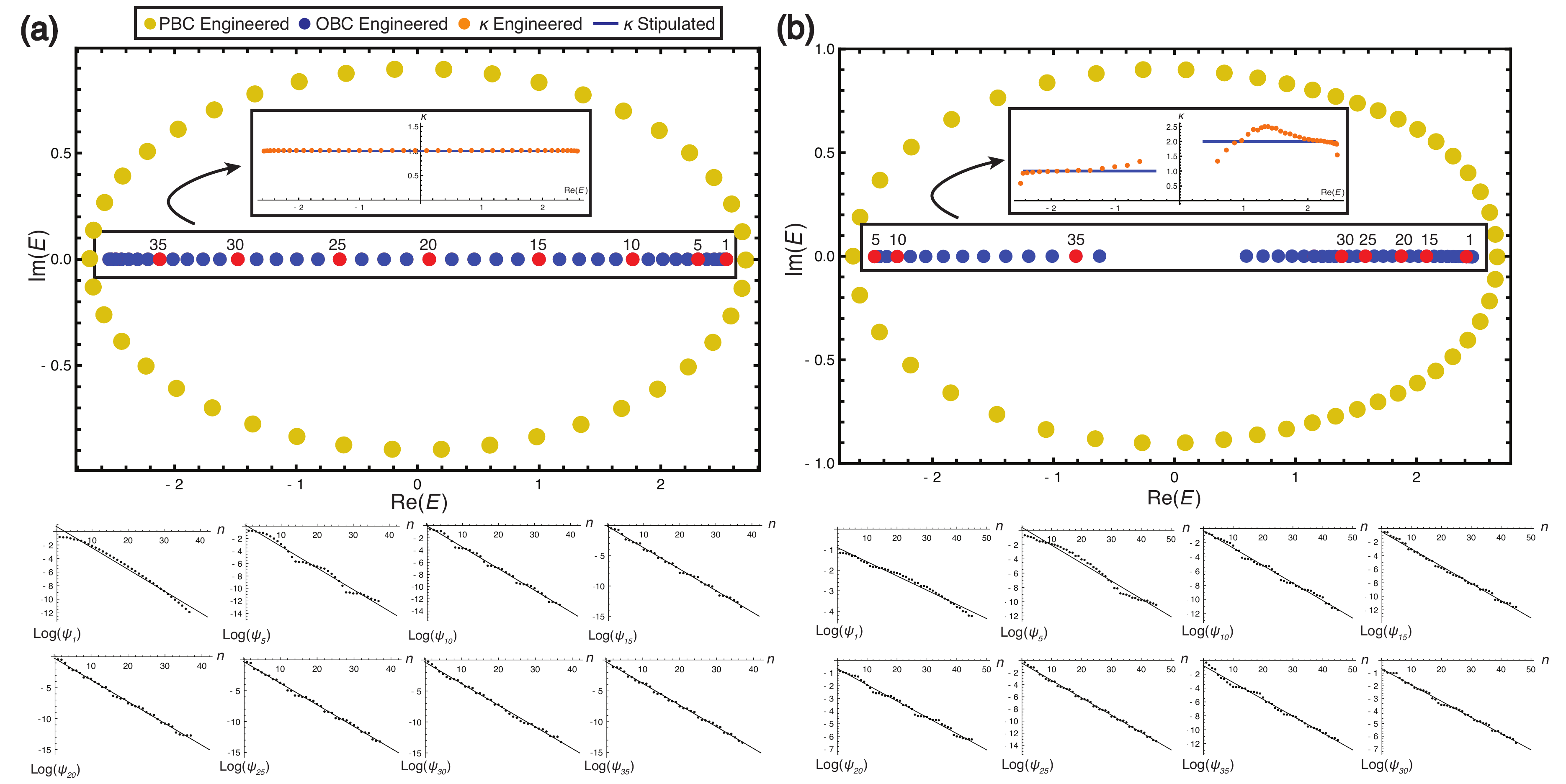}
\caption{The agreement of $\kappa(E)$ i.e. $V(x)$ stipulated and engineered profiles. \textbf{a} shows the Hatano-Nelson model complex bandstructure. Labelled site numbers corresponding to the eigenvalues marked in red with their respective eigenvectors plotted over every site shown below. The log of each eigenvalue was fitted to a straight line whose gradient is $\kappa$. Within the bandstructure, the value for $\kappa$ at each site was then plotted across the $\textrm{Re}(E)$. \textbf{b} A similar procedure was performed for the case of two disjointed OBC lines (Fig. 3a of the main text) with potentials set at $V=1$ for the leftward line and $V=1.6$ for the rightward line. The lack of 2-fold rotational symmetry gives rise to the $\kappa$ profile observed. However, averaged across the line, the value of $\kappa$ across the rightward line is twice the value as the leftward line. }
		\label{fig:S10}
	\end{figure*}
	
	In this section, we seek to verify that the stipulated engineered skin depth distribution $\kappa(E)$ along the OBC line indeed agree. It is known from the non-Hermitian Skin Effect that each state will decay exponentially from the system edge. In other words, pumping towards the edge causes an exponential accumulation of the edge state. By expressing the Hamiltonian in position space, one may obtain the eigenvalues and just as importantly the eigenstates in real space, where their spatial profiles can be compared with the stipulated $\kappa$ decay.
	
	We systematically log-plot the eigenvectors arranged by site and obtain the exponential decay factor $\kappa$. For the Hatano-Nelson warm-up model, the decay factor remains constant across all sites, which corresponds to the original stipulated potential $V=1$ in the electrostatic model up to normalization. This is expected since the Hatano-Nelson model is known to have a constant $\kappa$. Doing the same for the model (Fig. 3a of the main text) with two disjointed OBC lines with spatially inhomogeneous potential i.e. $\kappa$ profile, we find that for $V=1$ and $V=1.6$ respectively that the average of $\kappa$ across each line also agrees up to normalization. However, the exact profile need not agree entirely and is not expected to do so given the system does not respect 2-fold rotational symmetry.	

\section{Details of engineered spectra in the phase diagram in Fig.~\ref{fig:3}c of the main text}

In Fig.~\ref{fig:3}c of the main text, we studied how reliably our approach can engineer local Hamiltonians with real spectra, in a setting with no manifest symmetry that encourages real spectra. We used a Reuleaux triangle as the PBC spectral locus, and stipulated the target OBC eigenstates to have a maximum $\kappa(E)$ amplitude given by $A$ and OBC locus to be a real line segment of length $d$, as given in Eq.~\ref{Vx} of the main text. 

Due to the requirement of sufficient locality, some minute discrepancy from a purely real spectrum is unavoiable. Plotted in Fig.~\ref{fig:3}c of the main text, which is reproduced here as Fig.~\ref{figS4}a, is the maximum value of $\text{Im}(E)$ in the parameter space of $A$ and $d$, for hoppings truncated beyond 5 sites. The plot is helpful in searching for a parameter region in which the energies were purely real and thereby admitting experimentally realizable states. 
9
We do this by computationally sweeping a grid consisting of $A \in (-1,4)$ and $d \in (0,5.5)$ while setting the left most point of the OBC loci to be at $(-1,0)$, the choice of which was somewhat arbitrary. In Fig.~\ref{figS4}a we see a region in which the maximum magnitude of $\text{Im}(E)$ is very close to 0. Each point within the parameter space corresponds to a specific spectral configuration. A desirable configuration is already given in Fig.~\ref{fig:3}b of the main text ; some examples in which the imaginary eigenenergies are slightly non-vanishing are listed in Fig \ref{figS4}\textbf{b-f}. Here we have arbitrarily cut off the number of Fourier modes (coupling distance) at 5 and it is expected that as we increase the Fourier modes that the PBC spectrum will come closer to the stipulated PBC spectral loci.  

\section{Circuit Realization of non-Hermitian lattices}\label{circuits}

\begin{figure*}
		\includegraphics[width=\linewidth]{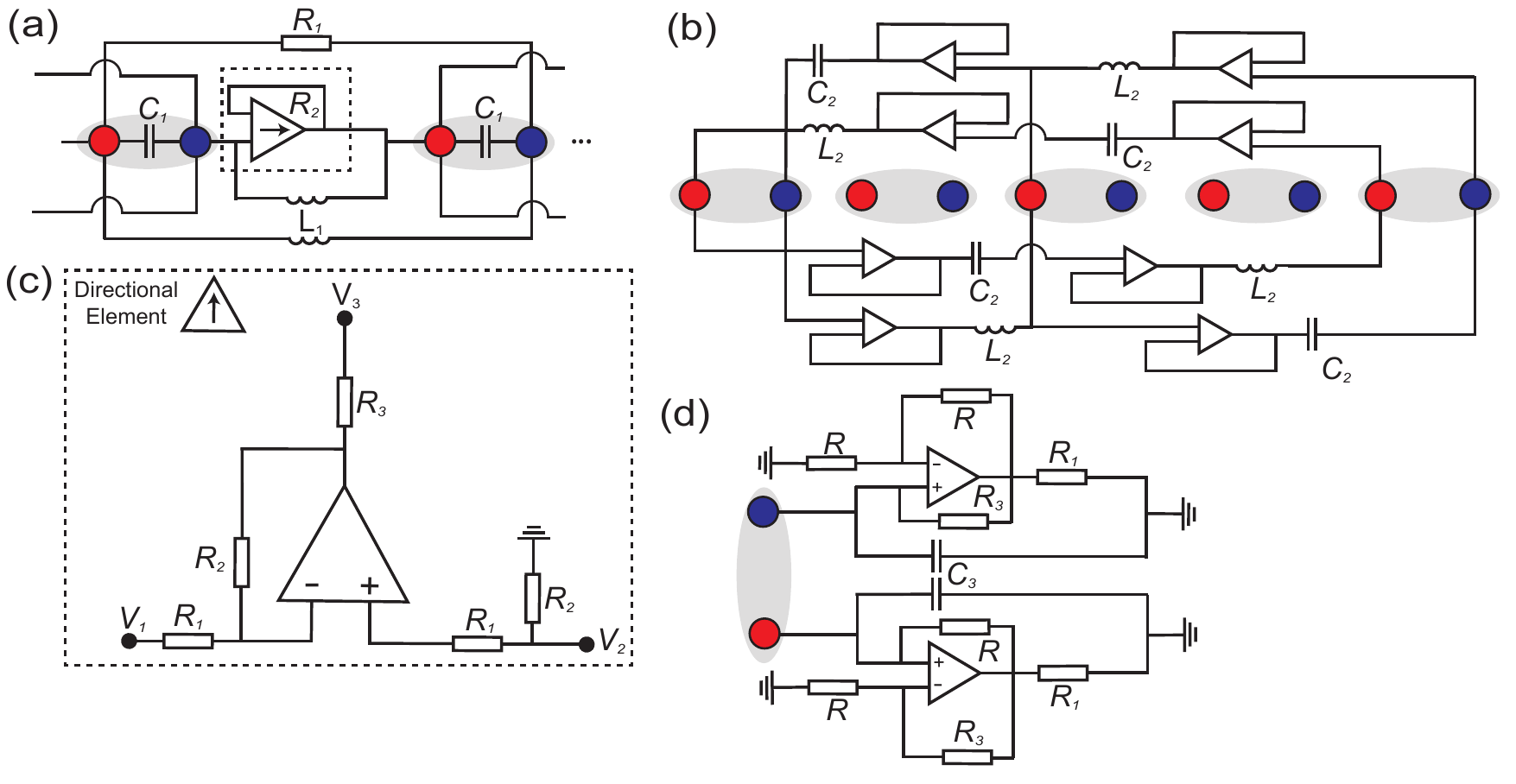}
\caption{\Rev{\textbf{a.} Generalized circuit realization of any 2-band, nearest neighbour hopping and \textbf{b.} next-nearest neighbour circuit. \textbf{c.} Schematic representation of the non-reciprocal element that leads to non-hermiticity using Negative Impedance Converters (INICs). \textbf{d.} Grounded nodes with each unit cell attached to active INIC elements effectively shift the total energy of the system, serving as an enabler for the realization of trace terms introduced in Sec \ref{subsubsection: FS Approx}.}}
		\label{fig:circuit}
\end{figure*}

\Rev{Topolectrical circuits~\cite{Lee2018_qf,Wu2020_mx,jalil_2020,Hofmann2020_ys,Helbig2020_yg,Hofmann2019_tz,Helbig2019_io} have provided a fresh avenue into realizing new phenomena. Their extreme versatility in the implementation of distant couplings, as well as precise control over linearity and/or non-linearity, has brought to reality many models that are otherwise hard to replicate in conventional materials or metamaterials. In this section, we present a generalized circuit implementation for realizing models containing up to next-nearest neighbor hoppings, including many of our designed models.}

\Rev{Any electrical circuit can be represented as a graph. Such a representation can be described by Kirchhoff's laws in nodal form \cite{Kagan2017,Cernova2017}.
\begin{equation}
\label{eqn:Kirchoff's Law}
\sum\limits_{j=1,j\neq i}^{N} c_{\rm{ij}}(V_{\rm{i}}-V_{\rm{j}}) = I_{\rm{i}}
\end{equation}}
\Rev{Here, $c_{ij}$ represent the admittances of the circuit elements between nodes $i$ and $j$, $V_i$ and $V_j$ are the electric potentials of nodes $i$ and $j$, and $I_i$ is the current flowing into the node from the sources connected directly to node $i$. } \Rev{In compact matrix notation, Kirchhoff's law for a list of N nodes can be expressed as}
\Rev{\begin{equation}
\label{eqn:Kirchoff's Law Matrix Form}
\bf{L}\bf{V}=\bf{I}
\end{equation}}
\Rev{where $\bf{L}$ is the circuit Laplacian. Written explicitly in terms of the matrix elements, the equation reads
\begin{equation}
\label{eqn:Explicit Circuit Laplacian}
\begin{pmatrix}
    c_{\rm{aa}} & c_{\rm{ab}} & \dots  & c_{\rm{aN}} \\
    c_{\rm{ab}} & c_{\rm{bb}} & \dots  & c_{\rm{bN}} \\
    \vdots & \vdots & \ddots & \vdots \\
    c_{\rm{Na}} & c_{\rm{Nb}} & \dots  & c_{\rm{NN}}
\end{pmatrix}
\begin{pmatrix}
   V_{\rm{a}}\\
   V_{\rm{b}}\\
   \vdots \\
   V_{\rm{N}}
\end{pmatrix}
=
\begin{pmatrix}
   I_{\rm{a}}\\
   I_{\rm{b}}\\
   \vdots \\
   I_{\rm{N}}
\end{pmatrix}
\end{equation}}
\Rev{Notice that if one strategically matches the admittances of each node in the Circuit Laplacian with numerical values that correspond to the original Hamiltonian, that we have effectively reproduced the Hamiltonian, albeit with a different physical interpretation as a current-voltage Laplacian. Most often, such circuits are implemented with capacitors and inductors with admittances (impedances) that depend on the frequency of a driving alternating current. Furthermore, non-hermiticity can be achieved with non-reciprocal elements that asymmetrically control the current flow in one direction differently from the other. In practice, this can be implemented with externally powered directional operational amplifiers whose amplification ratio can be finely tuned by the user. With reference to \cite{thomale_2021}, the circuit Laplacian reads
\begin{eqnarray}
L_{11}/i&=&L_{22}/i=\omega C_1 - \frac{2}{\omega L_1} - \frac{2}{\omega L_2} - \frac{1}{L_1} - \frac{i}{R_1} - \frac{i}{R_2}
\nonumber\\
L_{12}/i&=&-\omega C_1 + \frac{i}{R_1} e^{i k_1} - \frac{1}{\omega L_1} e^{i k_1} + \frac{i}{R_2} e^{- i k_1} - \frac{1}{\omega L_1} e^{-i k_1}
\nonumber\\
L_{21}/i&=&-\omega C_1 + \frac{i}{R_1} e^{-i k_1} - \frac{1}{\omega L_1} e^{-i k_1} + \frac{i}{R_2} e^{i k_1} - \frac{1}{\omega L_1} e^{i k_1}\nonumber\\
\label{eqn:Laplacian}
\end{eqnarray}}

\Rev{In order to realize for instance the Hamiltonian in Eqn. \ref{eqn:9}, one may simply tune the experimental values of the electrical components in Eqn.~\ref{eqn:Laplacian} to match the coefficients of the engineered Hamiltonian circuit with experimental components. In an  engineered Hamiltonian where a trace ansatz was used, a grounded node such as the one in Fig. \ref{fig:circuit}c can be used to set a trace term of choice}

\Rev{To measure the Laplacian eigenspectrum, one can do so from the impedance readings between all pairs of nodes, effectively measuring the circuit Laplacian node by node. The impedance $Z_{a0}$ between two nodes $a$ and a ground is given by 
\begin{equation}
\label{eqn:Symmetric Impedance}
Z_{\rm{a0}}=\frac{V_a}{I_{0}}=\sum\limits_{i} \frac{\psi_{a,i}\phi^*_{a,i}}{\lambda_i}
\end{equation}
where $\psi_{a,i}$ and $\phi_{a,i}$ are the left and right eigenvectors of the Laplacian respectively and $\lambda_i$ the eigenvalues of the Laplacian, which can then be solved by inverting the data obtained.}

\section{Additional exmaples}\label{additionalexamples}
\begin{figure}
\begin{minipage}{\linewidth}
\subfloat[]{\includegraphics[width=.46\linewidth]{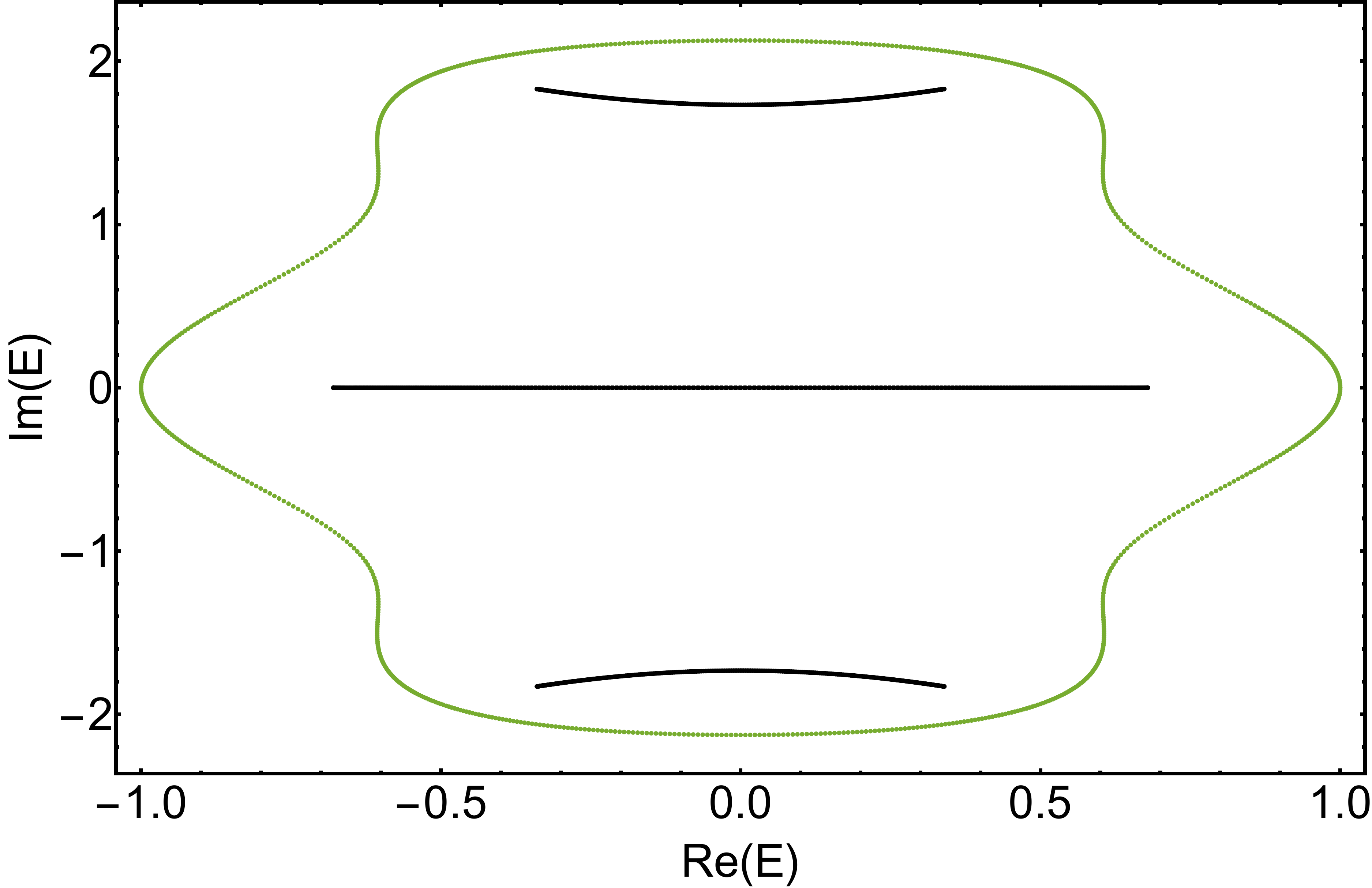}}
\subfloat[]{\includegraphics[width=.32\linewidth]{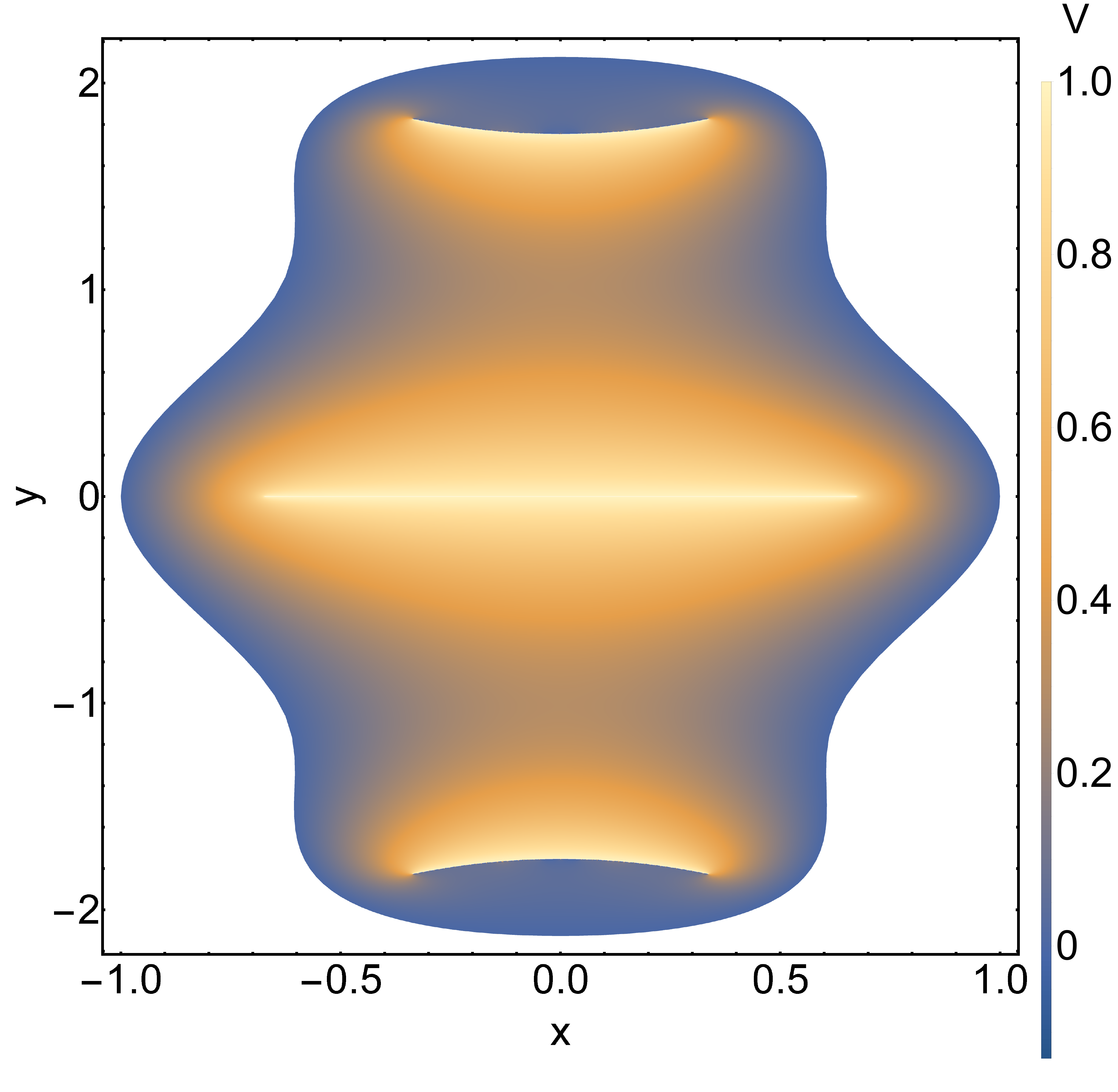}}
\subfloat[]{\includegraphics[width=.22\linewidth]{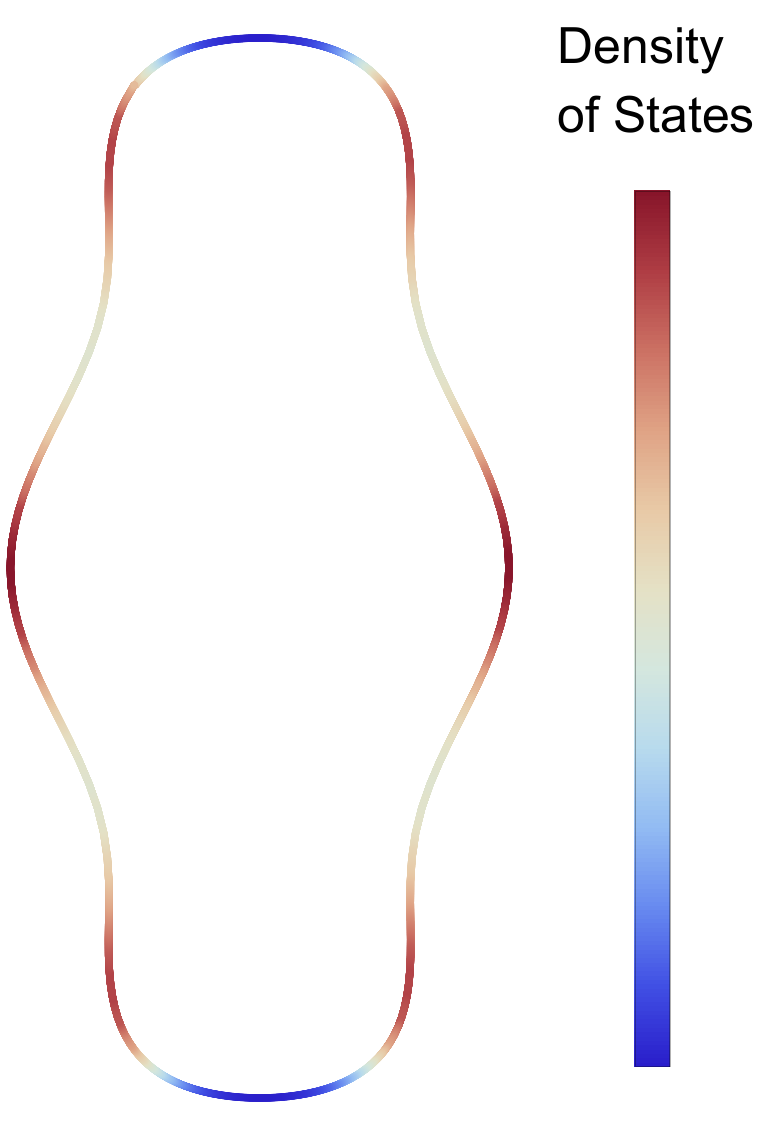}}
\end{minipage}
\caption{\Rev{(a) PBC and OBC spectra of the one-dimensional nonreciprocal superlattice model~\cite{PhysRevB.101.125418} with the parameters $\alpha=1/3,\gamma=2,\delta=0.6\pi,t=1$. (b) The corresponding solution to the Laplace equation within the stipulated domain. The OBC lines are set to a constant potential of 1 while the PBC locus is grounded.
(c) The resulting density of states along the PBC locus. On the y=0 line, the density of states is high due to close proximity to the OBC line, while the density of states is low at the top and bottom.}}
\label{fig:superlattice}
\end{figure}

\Rev{Our electrostatics formalism goes beyond examples with real OBC spectra, and can readily inverse engineer a complex spectra. For concreteness, we consider the one-dimensional nonreciprocal superlattice model proposed by Zeng et al~\cite{PhysRevB.101.125418}, where the nonreciprocal hopping is modulated by a cosinusoidal function. By tuning the modulation of the hoppings, they demonstrate robust real energy OBC spectra. Here, we would like to reverse engineer the Hamiltonian parameters given the spectra and the form of the Hamiltonian. We choose the parameters $\alpha=1/3,\gamma=2,\delta=0.6\pi,t=1$, which yields a periodic chain with 3 sites in each unit cell, i.e. a 3-band model. The OBC spectrum consists of a finite real line and two parabolas, while the shape of the PBC spectrum is more sophisticated than our warm-up examples. The OBC and PBC spectra are given in Fig.~\ref{fig:superlattice}. }

\Rev{To reiterate, suppose the parameters $\gamma$ and $\delta$ are not known a priori, we seek to find these 2 parameters given only the spectra and the form of the bulk momentum space Hamiltonian, given by}

\Rev{\begin{equation}
H(k) = t\begin{pmatrix}0&1+\gamma_1&(1-\gamma_3)e^{ik}\\1-\gamma_1&0&1+\gamma_2\\(1+\gamma_3)e^{-ik}&1-\gamma_2&0\\\end{pmatrix}
\end{equation}}
\Rev{where $\gamma_j=\gamma\cos(2\pi\alpha j+\delta)$ is the modulation. By setting $t=1$ without loss of generality, we yield the characteristic equation}

\Rev{\begin{equation}
-E^3 + c~E = g~e^{ik} + h~e^{-ik}
\end{equation}}

\Rev{
where $c = 3-\gamma_1^2-\gamma_2^2-\gamma_3^2$, $g = -(1-\gamma_1)(1-\gamma_2)(1-\gamma_3)$ and $h=-(1+\gamma_1)(1+\gamma_2)(1+\gamma_3)$. Given the numerical values of $g$ and $h$, we may obtain $\gamma$ and $\delta$ via a root finding algorithm. Thus, we will focus only on determining $g$ and $h$. Note that by taking the real (imaginary) parts on both sides of the characteristic equation, we end up with a sinusoid of amplitude $g+h$ ($|g-h|$) on the right-hand side. Thus, to find the values of $g$ and $h$, we must find $E(k)$ for one energy band (any suffice) and determine the numerical value of the constant $c$ such that $-E^3+cE$ is purely sinusoidal. After this value of $c$ is determined, the maximum value f $\Re[-E^3+cE]$ can be computed; this is $g+h$. The maximum value of $\Im[-E^3+cE]$ is similarly $g-h$.}

\Rev{To find the energy dispersion of a single band, we follow the same procedure as described in our paper (Fig.~\ref{fig:S1} Steps 1 to 4). In our formalism, we prescribe that the entire OBC line has a constant skin depth which we can set to $V=1$ without loss of generality. The PBC line is grounded. The resulting solution of the Laplace Equation is Fig.~\ref{fig:superlattice}b.
}

\Rev{As mentioned, the density of states (Fig~\ref{fig:superlattice}c) is proportional to the charge density. The charge density is much higher on the $y=0$ line due to the close proximity to the OBC line (and hence a higher $\frac{\partial E}{\partial x}$). From the density of states, we can then assign a region of 1/3 the length of the PBC line to be a single energy band, from which we can compute the values of the constants $g$ and $h$. To clarify, Step 5 in Fig~\ref{fig:S1} is only done to construct a Hamiltonian that recovers the same OBC and PBC spectra, which is typically not unique. In this example, our ansatz is the form of the Hamiltonian itself $H(k)$ and we seek to recover the parameters $\gamma$ and $\delta$.}

\Rev{In general, our method could be readily applied to high-dimensional systems where the Hamiltonian is separable. One example is the 2D Hatano-Nelson model on a square lattice with asymmetric left/right and up/down hopping amplitudes. For generic higher-dimensional systems, we could formulate the system as an array of 1D chains, again with asymmetric hopping amplitudes.
}

\end{document}